\documentclass{aa}
\usepackage{newtxmath}
\usepackage{newtxtext}
\usepackage{natbib}
\usepackage{graphicx}
\usepackage{cuted}
\usepackage{capt-of}
\usepackage{dblfloatfix}
\usepackage{threeparttable}
\usepackage{amsmath}
\usepackage[caption=false]{subfig}
\usepackage{upgreek}
\usepackage{lscape}
\usepackage{fix-cm}
\usepackage[colorlinks,allcolors=blue]{hyperref}

\begin{document} 
\title{SiO emission in the filamentary Infrared Dark Cloud G035.39-00.33: An ALMA view}
\titlerunning{SiO emission in IRDC G035.39-00.33}
\authorrunning{R. Liu et al.}
\author{Rong Liu\inst{1,2}, Izaskun Jiménez-Serra\inst{3}, Giuliana Cosentino\inst{4,5}, Jonathan C. Tan\inst{6,7}, Ashley Thomas Barnes\inst{4}, Francesco Fontani\inst{8}, Paola Caselli\inst{9}, Antonio Martínez-Henares\inst{3}, Chi-Yan Law\inst{8}, Jonathan D. Henshaw\inst{10}, Tie Liu\inst{2}}
\institute{National Astronomical Observatories of China, Chinese Academy of Sciences, Beijing, 100012, China\\
\email{liu\_rong@bao.ac.cn} 
\and
Shanghai Astronomical Observatory, Chinese Academy of Sciences, 80 Nandan Road, Shanghai, 200030, China\\
\and
Centro de Astrobiología (CAB), CSIC-INTA, Carretera de Ajalvir km 4, 28850 Torrejón de Ardoz, Spain
\and
Institute de Radioastronomie Millimétrique, 300 Rue de la Piscine, 38406 St-Martin-d’Hères, France 
\and
European Southern Observatory (ESO), Karl-Schwarzschild-Straße 2, 85748 Garching bei München, Germany
\and
Department of Space, Earth and Environment, Chalmers University of Technology, SE-412 96 Gothenburg, Sweden
\and
Department of Astronomy, University of Virginia, 530 MeCormick Road Charlottesville, 22904-4325 USA
\and
INAF Osservatorio Astronomico di Arcetri, Largo E. Fermi 5, 50125 Florence, Italy
\and
Max Planck Institute for Extraterrestrial Physics, Gießenbachstraßse 1, D-85748 Garching bei München, Germany
\and
Astrophysics Research Institute, Liverpool John Moores University, 146 Brownlow Hill, Liverpool L3 5RE, UK
}
\date{Received 2025 August 1; accepted 2025 December 21}
  \abstract 
   {}
   {Filamentary infrared dark clouds (IRDCs) are believed to represent the initial conditions for massive star and cluster formation.}
   {We investigate the IRDC G035.39-00.33 using SiO, H$^{13}$CO$^{+}$, CH$_3$OH, and CS emission observed with ALMA at  3.5\arcsec\ resolution ($\sim$0.05 pc). 
   The analysis of the SiO emission provides a record of shock activity within the cloud, offering insights into both the current level of star formation and the cloud's formation mechanisms.}
   {We identify several regions with broad SiO emission clearly associated with outflows, pinpointing the locations of ongoing star formation across the cloud.
   The ALMA images also reveal a series of spatially extended SiO emission spots with narrow line profiles, aligned along an arc-like path that is also seen in CS and CH$_3$OH emission. While the broad SiO emission is mainly associated with the main cloud filament, as seen in visual extinction, the narrow SiO arch is located at the edge of the cloud, far from the identified sites of star formation activity. 
   The presence of these arc-like morphologies suggests that large-scale shocks may have compressed the gas in the surroundings of the G035.39-00.33 cloud, shaping its filamentary structure. 
   By inspecting the large-scale radio continuum emission around G035.39-00.33, we find that this IRDC is part of a larger star-forming complex where the densest and coolest material appears at the interacting regions between a Supernova Remnant (SNR) and an expanding H\textsc{ii} region. In particular, we hypothesize that this IRDC may be spatially coincident with the ionized expanding gas associated with the previously identified SNR G35.6-0.4.}
   {We suggest that collisions between giant molecular clouds and expanding gas flows from interacting SNRs and H\textsc{ii} regions may be responsible for the observed arc-like structures. Such shock compressions could play an important role in the formation of IRDCs and in the potential triggering of star formation.}
   \maketitle
\nolinenumbers 
\section{Introduction}
\label{sec1}
Giant molecular clouds (GMCs) are the largest reservoirs of molecular gas in galaxies, with masses above 10$^4$$M_{\odot}$ \citep{Chevance2021AAS}. They provide the primary sites for star formation, particularly for massive star formation \citep{McKee2007}.
Within GMCs, infrared dark clouds (IRDCs) are the coldest and densest regions, often regarded as potential precursors of high mass star-forming regions \citep{Rathborne2006}. They are characterized by their low temperatures \citep[$T_d$ < 25 K; e.g.][]{Pillai2006A&A, Peretto2010A&A, Ragan2011ApJ}, massive reservoirs of molecular gas ($\sim$ 10$^2$ - 10$^5$$M_{\odot}$), and high column densities \citep[N(H$_2$)$\ge$10$^{22}$ cm$^{-2}$;][]{Rathborne2006}. 
IRDCs were first discovered by the \textit{Infrared Space Observatory (ISO)} as dark silhouettes against the diffuse mid-infrared (mid-IR) Galactic background \citep{Perault1996}. These clouds exhibit a broad range of morphologies, from elongated filaments to more compact globular shapes \citep{vanLoo2007}. 
Thanks to their large reservoir of dense and cold material, IRDCs have the potential to host the formation of even massive stars and stellar clusters \citep{Rathborne2006, Molinari2010PASP, Sokolov2018A&A}. 
Infrared and sub-millimetre observations have subsequently revealed that IRDCs fragment into dense and massive cores, displaying a variety of star formation stages. Prestellar cores remain dark and cold, devoid of any internal heating sources \citep{Zhang2007A&A, Rathborne2008ApJ}. In contrast, active cores are infrared-bright and chemically enriched, often exhibiting clear signatures of star formation, such as embedded protostars and molecular outflows driven by accreting young stellar objects (YSOs) \citep{Pillai2006A&A, Sanhueza2012ApJ, Tan2014prpl.conf, Wang2014MNRAS, Sokolov2017A&A, Moser2020ApJ}.
Hence, IRDCs are thought to host the pristine conditions from which stars form. However, the mechanisms responsible for initiating star formation in these objects are still under debate.

Current scenarios suggest that star formation in IRDCs is either triggered during the cloud formation process itself or driven by external feedback such as nearby H\textsc{ii} regions or Supernova Remnants (SNRs).
In the flow-driven model, molecular clouds are the result of the collision of high-velocity, diffuse atomic gas flows \citep{Ballesteros-Paredes1999, Heitsch2009}. 
In the shock-induced scenario, pre-existing molecular clouds are compressed by propagating large-scale shocks. These shocks, with high Mach numbers, primarily transmit momentum and pressure, rapidly compressing the gas and triggering fragmentation and subsequent star formation \citep{Koyama2000, Koyama2002, vanLoo2007, VanLoo2013ApJ, Wang2014MNRAS}. Such shocks can arise from stellar feedback \citep{Inutsuka2015A&A}, cloud-cloud collisions \citep{Tasker&Tan2009ApJ}, and galaxy interactions \citep{Appleton2006ApJ}. While both mechanisms involve converging gas motions, flow-driven accumulation proceeds gradually on global scales, whereas shock compression represents a rapid, localized process that leaves distinct kinematic and chemical imprints.

From an observational point of view, large-scale shocks are expected to leave an imprint of the gas dynamics and kinematics within the impacted cloud. 
A unique molecular tracer of this imprint is Silicon Monoxide (SiO). This molecule is highly depleted in the quiescent gas of molecular clouds, but it is observed to be enhanced by several orders of magnitude in regions affected by shocks \citep{martin1992sio, 2005Jimnez, jimenez2010parsec, Cosentino2019}. SiO emission is typically observed in regions where shocks destroy and/or erode dust grains, releasing silicon atoms into the gas phase and yielding SiO \citep{schilke1997sio, 1997Caselli, 2008Jimnez}.
The SiO line emission often presents either a broad velocity profile and/or a narrow velocity profile. The broad components, typically presenting a compact morphology in IRDCs, reflect the kinematic signatures of powerful outflows from protostellar objects \citep{qiu2007high, jimenez2010parsec, duarte2014sio, Rong2022}. In contrast, the narrow SiO emission typically shows a more extended distribution, even over parsec scales \citep{jimenez2010parsec, louvet2016tracing, Cosentino2020}. This emission has been interpreted either as decelerated entrained material \citep{lefloch1998widespread}, or as a product of large-scale shocks due to cloud-cloud collisions and/or stellar feedback in the form of H{\small II} regions and SNRs  \citep{Cosentino2019, Cosentino2020, Cosentino2022, Rong2025, Cosentino2025A&A}.

\begin{figure*}[htbp]
    \centering
    \includegraphics[width=1\linewidth]{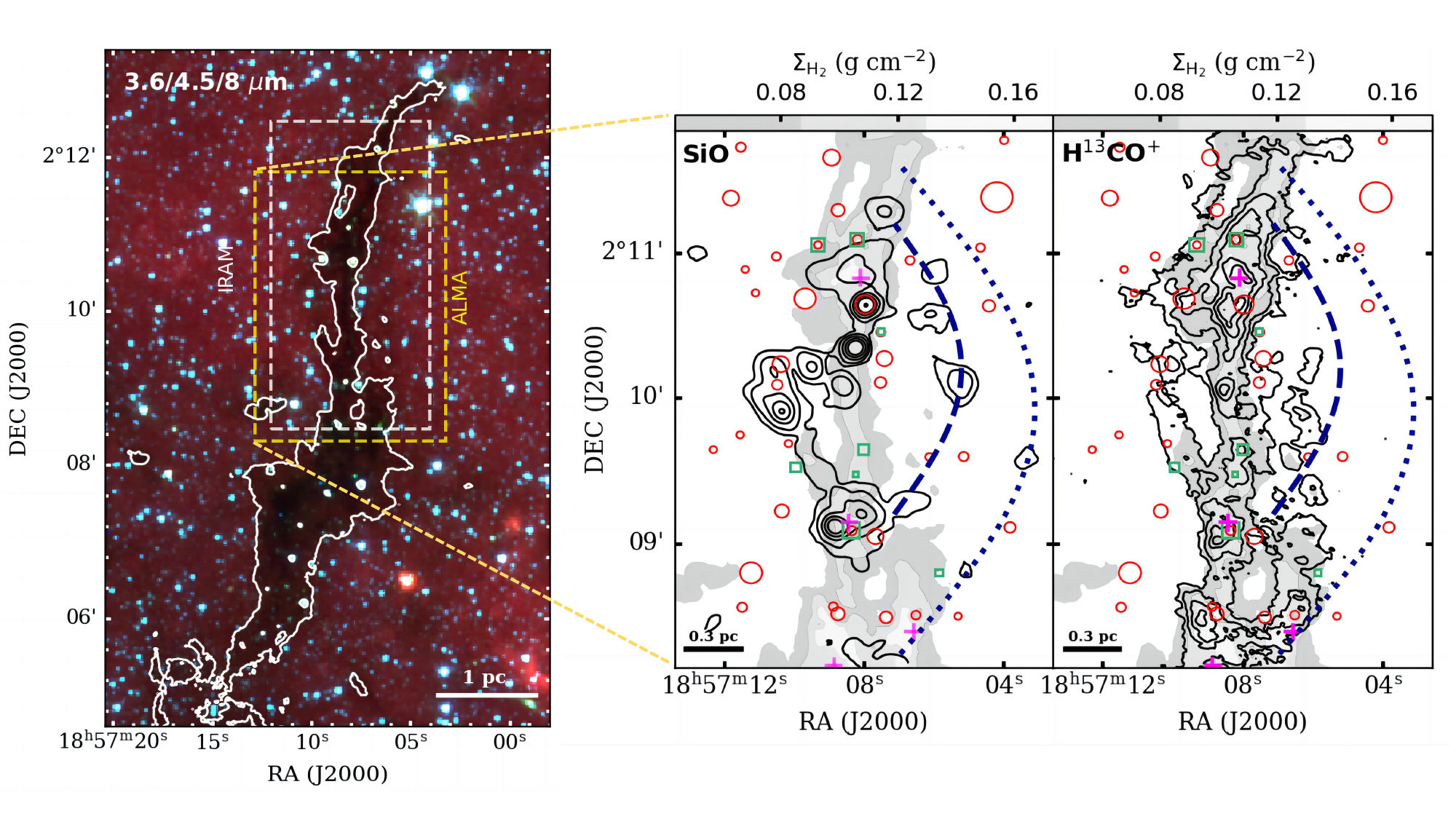}
\caption{{\it Left panel}: Three-color image (red color = 8 $\upmu$m, green color = 4.5 $\upmu$m, blue = 3.6 $\upmu$m) of Cloud H, obtained from GLIMPSE \citep{benjamin2003, Carey2009}. The white contour highlights the dense material in the IRDC as shown by the mass surface extinction map of \citet{Kainulainen&Tan2013}, corresponding to an extinction level of $A_{\rm V}$= 10 mag. The dashed yellow and white rectangles are the fields of view for the ALMA and IRAM-30 m observations, respectively. The physical scale is shown in the lower-right corner.
{\it Right panels}: The grey background colour scale shows the mass surface density, with contour levels of $A_{\rm V}$ = 10, 20, 30, and 40 mag. The black contours represent SiO (2-1) and H$^{13}$CO$^+$ (1-0) emission from ALMA observations. The SiO data are spatially smoothed to a resolution of 6$^{\prime\prime}$, and the spectral resolution was degraded from 0.21 to 0.6 km s$^{-1}$. The SiO contours are integrated over the velocity range from -20 to 120 km s$^{-1}$, and levels are 21 (3$\sigma$), 210, 630, 1050, and 2300 (peak value) mJy beam$^{-1}$ km s$^{-1}$. The H$^{13}$CO$^+$ contours are integrated over the velocity range from 40 to 50 km s$^{-1}$, with the levels of 15 (3$\sigma$), 50, 100, 151, 200, 253 (peak value) mJy beam$^{-1}$ km s$^{-1}$. Red open circles and medium green rectangles indicate the positions of 8 $\upmu$m sources and 24 $\upmu$m sources, respectively \citep{jimenez2010parsec}. The marker sizes for the 8 $\upmu$m and 24 $\upmu$m sources are scaled by the source flux. Magenta crosses mark the locations of the cores identified by \citet{Rathborne2006} using 1.2 mm continuum emission. Dark blue dashed and dotted lines trace the arc-like structures observed in the SiO emission. The physical scale is shown in the lower-left corner of each panel.}
    \label{fig1}
\end{figure*}

Examples of spectrally broad (narrow) and spatially compact (extended) SiO emission have been reported in the literature \citep{jimenez2010parsec, duarte2014sio, louvet2016tracing, Cosentino2020}. 
Among these, \citet{jimenez2010parsec} used IRAM-30m observations of SiO (2-1) and SiO (3-2) to investigate the morphology and kinematics of shocked gas in the IRDC G035.39-00.33 \citep[hereafter Cloud H, as first reported by][]{Butler&Tan2009}. 
Cloud H is located at a distance of 2.9 kpc \citep{Simon2006}, with a total mass of $\sim$16700 $M_{\odot}$, derived from combined near-IR (NIR) and mid-IR (MIR) extinction maps \citep{Kainulainen&Tan2013}. 
Toward this source, \citet{jimenez2010parsec} reported not only broad SiO emission localized in the dense regions, indicating ongoing star formation, but also faint, widespread SiO emission on parsec scales, suggesting that this emission could be due to large-scale shock interactions induced by cloud-cloud collisions. Subsequent studies revealed widespread CO depletion \citep[$f\rm{_{D}}$$\sim$5;][]{Hernandez2011II} and a high deuterium fractionation (D$\rm{_{N_2H^+}}$) of N$_2$H$^+$ \citep[mean D$\rm{_{N_2H^+}}$ = 0.04;][]{Barnes2016} in Cloud H, indicating that Cloud H is chemically evolved but has been influenced by less stellar feedback.
Furthermore, the dust temperature exhibits a decrease from the edge of Cloud H toward its inner regions, where the most massive protostellar cores are located, with no significant extended heating from ongoing star formation \citep{Nguyen2011A&A}. However, observations of NH$_3$ have shown that the kinetic gas temperature is generally elevated in the vicinity of massive protostellar cores, suggesting localized heating due to embedded star formation activity \citep{Sokolov2017A&A}. 

The kinematics of Cloud H reveal the merging of three filaments, each characterized by different velocity components: Filament 1 (42$\sim$44 km s$^{-1}$), Filament 2 (44$\sim$46 km s$^{-1}$, and Filament 3 (46$\sim$48 km s$^{-1}$) \citep{Henshaw2013, jimenez2014, Henshaw2014}. High-resolution observations further indicate that Cloud H is undergoing a multi-scale fragmentation process, where interactions between velocity-coherent narrow fibers ($\sim$0.028 pc) are observed \citep{Henshaw2016, Henshaw2017}. This fragmentation behavior suggests a strong influence of magnetic fields, as further supported by \citep{Liu2018ApJ}. Additionally, \citet{jimenez2010parsec, Liu2018ApJ, Bisbas2018MNRAS} propose that Cloud H may have formed via cloud-cloud collision (CCC). However, due to the lack of high-angular resolution observations, the origin of the SiO emission remains uncertain, making it challenging to establish whether it is truly associated with a large-scale shock interaction or with a population of deeply embedded low-mass protostars. 

In this paper, we present new high-angular resolution images of the SiO emission obtained with ALMA toward Cloud H. These images resolve the compact broad SiO emission and the extended narrow SiO component across the cloud, providing further insights into the formation mechanism of Cloud H. The paper is organized as follows: Details of the ALMA observations are provided in Sect.~\ref{sec2}, followed by the main results in Sect.~\ref{sec3}. The possible formation scenario of Cloud H is discussed in Sect.~\ref{sec4}, and a summary of our findings is presented in Sect.~\ref{sec5}.

\begin{figure*}
\vspace{4pt}
\centerline{\includegraphics[width=1\linewidth]{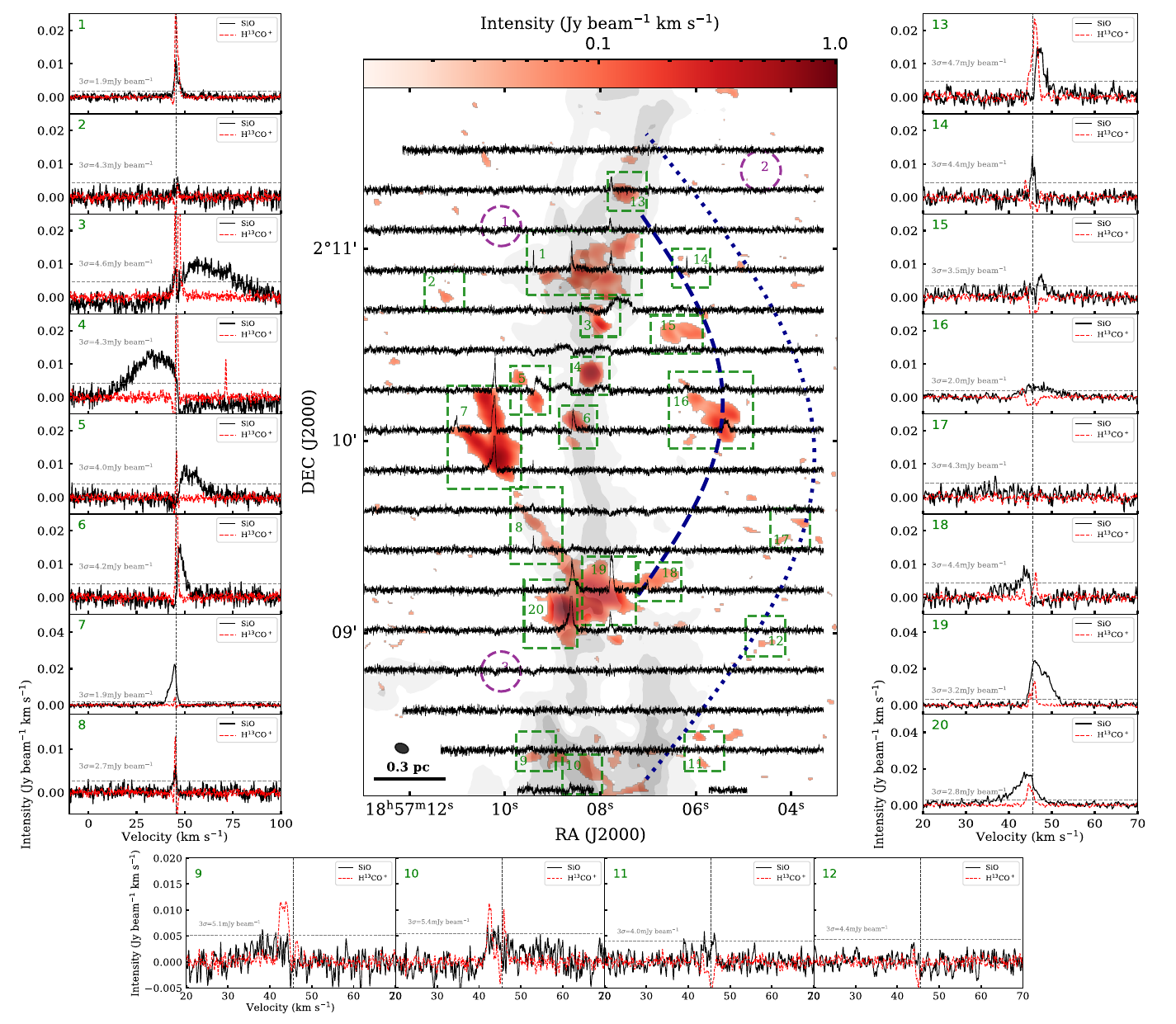}}
\caption{The middle upper panel shows the SiO (2-1) integrated intensity map, with the SiO spectral grid overlaid. The grid size corresponds to 12$^{\prime\prime}\times12^{\prime\prime}$. Gray contours represent the dense material, with contour levels of $A_{\rm V}$ = 10, 20, 30, and 40 mag. Dark blue dashed and dotted lines trace the arc-like structures. Green dashed rectangles indicate the regions from which the SiO and H$^{13}$CO$^+$ spectra are extracted (shown in the left, middle lower, and right panels). The beam size (4.32$^{\prime\prime}\times$2.93$^{\prime\prime}$) and the physical scale are shown in the lower-left corner. The left, middle lower, and right panels show the averaged spectra of SiO (black lines) and H$^{13}$CO$^+$ emission (red lines) extracted from the green dashed regions in the middle upper panel. The H$^{13}$CO$^+$ intensities have been divided by a factor of two for clarity. The green numbers in the upper left corner of each panel correspond to the spectra extracted from the matching numbered green rectangles in the middle upper panel. The purple circles indicate regions with no detectable SiO emission.
The gray dashed parallel line in each spectrum indicates the 3rms noise levels of the SiO emission. The vertical dashed black line marks the central velocity of Cloud H (45.5 km s$^{-1}$).}
\label{fig2}
\end{figure*}

\section{Observations}
\label{sec2}
We utilized data from ALMA project 2016.1.01363.S (PI: I. Jiménez-Serra), which includes observations from both the Atacama Compact Array (ACA) and the 12-m array in the C40-1 configuration. 
The observations were conducted in Band 3 during Cycle 4, spanning December 2016, January 2017, and March 2017. The 12-m array baselines range from 15 m to 331 m, while the ACA baselines span from 8.9 m to 45 m.
The central coordinates of Cloud H are $\alpha$(J2000) = 18$^h$57$^m$08$^s$, $\delta$(J2000) = 02$^{\circ}$10$^{\prime}$30$^{\prime\prime}$ (l = 35.517$^{\circ}$, b = -0.274$^{\circ}$).  
The synthesized beam sizes for the ACA and 12-m array observations are 15.1$^{\prime\prime}$ and 3.5$^{\prime\prime}$, respectively.
The maximum recoverable scale (MRS) is 56.8$^{\prime\prime}$$\sim$76$^{\prime\prime}$ for the ACA, and 28.5$^{\prime\prime}$$\sim$30.6$^{\prime\prime}$ for the 12-m array.
The spectral line sensitivity per 0.061 MHz channel (equivalent to $\sim$0.2 km s$^{-1}$) is 0.027 Jy (0.36 K) for the ACA and 0.01 Jy (0.13 K) for the 12-m array.

The spectral setup includes seven spectral windows (SPWs) ranging from 86.31 to 99.39 GHz.
These windows covered the molecular lines of H$^{13}$CO$^+$ $J$=1-0, SiO $J$=2-1, CH$_{3}$OH $J$=2-1 and CS $J$=2-1. 
All data calibration and imaging were carried out using the CASA software package version 5.6 \citep{McMullin2007casa}. 
For data reduction, the ACA and 12-m array data were calibrated separately, with continuum subtraction performed using line-free channels. The two datasets were then combined using the concat task, and imaging was carried out with Briggs weighting (robust = 0.5). The final synthesized beam sizes for the SiO, H$^{13}$CO$^+$, CH$_{3}$OH, and CS data are 4.34$^{\prime\prime}$ $\times$ 2.93$^{\prime\prime}$, 4.35$^{\prime\prime}$ $\times$ 2.93$^{\prime\prime}$, 3.93$^{\prime\prime}$ $\times$ 2.63$^{\prime\prime}$, and 3.89$^{\prime\prime}$ $\times$ 2.59$^{\prime\prime}$, respectively.
To assess the missing flux in the ACA and 12-m combined data, we deconvolved the SiO images to match the resolution of the IRAM observations at 3$\,$mm and found that the ALMA SiO images recovered about 60\% of the total flux measured with the IRAM 30m$\,$ telescope. However, due to the limited sensitivity of the IRAM 30m data, they were not included in the final SiO combination. In contrast, for the CH$_3$OH data cube, the combination of IRAM 30m, ALMA 12m, and ACA data successfully recovered the majority of the flux.

\section{Results}
\label{sec3}
In Figure \ref{fig1}, the left panel shows the three-color map (red = 8 $\upmu$m, green = 4.5 $\upmu$m, blue = 3.6 $\upmu$m) obtained from \citet{benjamin2003, Carey2009}, with white contours highlighting the cloud structure and encompassing material denser than 0.044 g cm$^{-2}$ (A$_{\rm V}$=10 mag) in the IRDCs, as reported by \citet{Kainulainen&Tan2013}. 
The white- and yellow-dashed rectangles indicate the fields of view for the IRAM 30m \citep{jimenez2010parsec} and ALMA observations, respectively. 
Zooming in on the ALMA field of view (right panels), we present the SiO and H$^{13}$CO$^+$ emission as black contours, integrated over the velocity ranges of -20 – 120 km s$^{-1}$ and 40 – 50 km s$^{-1}$, respectively, which cover the main velocity components of the region. The SiO data were spatially smoothed to a resolution of 6$^{\prime\prime}$, and the spectral resolution was degraded from 0.21 to 0.6 km s$^{-1}$. Moment masking was then applied using the Behind The Spectrum (BTS) procedure to improve the noise level and ensure a more robust detection \citep{Clarke2018MNRAS}. The grayscale contours show the mass surface density and correspond to extinction levels of $A_{\rm V}$ = 10, 20, 30, and 40 mag. 
The flux lower limits for the 8 $\upmu$m sources (indicated by red open circles) and the 24 $\upmu$m sources (shown as medium green rectangles) are 3.5 mJy and 2 mJy, respectively, from \cite{jimenez2010parsec}.

\begin{table*}
\renewcommand\tabcolsep{3.5pt}
\centering 
\caption{Observed parameters of the SiO (2-1) line emission toward positions in Cloud H.}
\scalebox{0.9}{
\label{tab1}
\begin{tabular}{ccccccccccccccccc}
\hline
\hline
Position & $F$ & ${\upsilon}_{\rm LSR}$& $\Delta {\upsilon}$ & noise & Area & Position & $F$ & ${\upsilon}_{\rm LSR}$ & $\Delta {\upsilon}$ & noise & Area\\
& mJy beam$^{-1}$ & km s$^{-1}$   & km s$^{-1}$  & mJy beam$^{-1}$ & arcsec$^{2}$ & 
& mJy beam$^{-1}$ & km s$^{-1}$ &km s$^{-1}$ & mJy beam$^{-1}$  & arcsec$^{2}$ \\
\hline
1  & 6.9±0.2  & 45.3   & 1.2  & 0.6 & 36.0$\times$20.2   & 
11 & 2.7±0.9  & 45.1     & 3.9  & 1.3 & 12.5$\times$12.5 \\
 & 4.5±0.4  & 46.4     & 4.3  &  &          &
12 &    -     &     -    &  - & 1.5  & 12.5$\times$12.5          \\
2  & 3.3±0.9  & 45.7     & 3.5  & 1.5 & 12.5$\times$12.5 & 
13 & 14.7±0.8 & 47.3     & 2.2  & 1.6 & 12$\times$12     \\
3  & 9.7±2.7  & 62.4     & 28.5 & 1.5 & 12.5$\times$12.0   & 
14 & 10.1±0.5 & 45.6     & 1    & 1.5 & 12$\times$12     \\
4  & 13.2±2.1 & 32.4     & 19.5 & 1.4 & 12$\times$12     & 
15 & 3.5±0.9  & 47.1     & 6    & 1.2 & 16.3$\times$12.0   \\
5  & 7.7±1.4  & 53.5     & 10.2 & 1.3 & 12.5$\times$15.4 & 
16 & 3±0.7    & 47.1     & 8.7  & 0.7 & 26.4$\times$24.0   \\
6  & 11.7±0.6 & 47.1     & 1.9  & 1.4 & 12.0$\times$13.4   & 
17 & 1.7±2.1  & 36.7     & 19.4 & 1.4 & 12.5$\times$12.5 \\
  & 8.6±0.9  & 49.2     & 3.8  &   &           &
18 & 3.0±1.1    & 39.7     & 5.8  & 1.5 & 13.4$\times$12.0   \\
7  & 11.3±0.5 & 43.1     & 4.7  & 0.6 & 23.0$\times$32.2   &     
  & 7.4±0.7  & 43.8     & 2    &   &           \\
 & 14.5±0.3 & 44.8     & 1.7  &  &           & 19 & 17.6±0.5 & 45.9     & 2    & 1.1 & 16.8$\times$21.6 \\
8  & 4.6±0.4  & 44.9     & 2.1  & 0.9 & 16.3$\times$24.0   &        
 & 16.9±0.7 & 48.4     & 4.2  &   &           \\
9  & 3.2±1.6  & 39.1     & 7.8  & 1.7 & 12.5$\times$12.5 & 
20 & 4.4±1.2  & 41       & 14.1 & 0.9 & 16.8$\times$21.6 \\
10 & 4.2±1.3  & 43.9     & 4.7  & 1.8 & 12.5$\times$12.5 &            
& 12.9±0.6 & 44.1     & 4.1  &  & \\
\hline
\end{tabular}}
\end{table*}

In the middle panel of Figure \ref{fig1}, we observe a northern SiO emission associated with the MM7 core (marked by a magenta cross) and an 8 $\upmu$m source. The MM7 core, located within the densest part of the cloud, was identified by \citet{Rathborne2006} based on 1.2 mm continuum emission. 
Further south along the cloud, the SiO emission is associated with both 8 $\upmu$m and 24 $\upmu$m sources and spatially associated with the previously identified core MM6 \citep{Rathborne2006, Butler&Tan2009}. This southern SiO emission exhibits two clearly elongated tails: one extends toward north-east (with north being the direction of increasing declination and east being the direction of increasing Right Ascension), appearing adjacent to an unipolar outflow; while the other extends north-west, connecting to a weaker, discontinuous arc-shaped structure (inner arc) oriented from south to north, as indicated by dark blue dashed lines. 
At the bottom of the middle panel, we find an outer arc, which displays an arc-shaped morphology, also marked by dark blue dotted lines. Although this structure is located at the edge of the mosaic, its morphology is similar to the arc-shaped emission seen in CS toward the same part of the cloud (see Figure \ref{fig4}). Notably, both arcs lack associated 8 $\upmu$m and 24 $\upmu$m sources.
The H$^{13}$CO$^+$ emission in the right panel is tightly associated with the dense material in Cloud H and exhibits a bow-shaped structure extending toward the northeast at the bottom of the map, which is spatially similar to the SiO emission.

In Figure \ref{fig2}, we show the spatial distribution of SiO emission (red scale) along with the average spectra of SiO and H$^{13}$CO$^+$ extracted from 20 selected regions across the map. The H$^{13}$CO$^+$ intensities have been divided by a factor of two for clarity. These regions, labeled 1 to 20, are indicated by green dashed rectangles in Figure \ref{fig2}. The purple dashed circular regions (with a diameter of about 12$^{\prime\prime}$), labeled 1 to 3, mark the positions with no detectable SiO emission.
The size of each extracted rectangle area is listed in Table~\ref{tab1}, and all are larger than the beam size. 
By comparing the SiO line widths with the broadest H$^{13}$CO$^+$ linewidth (2.3 km s$^{-1}$), as done in \citet{Rong2025}, we find that most of the SiO spectra display complex line profiles. These include broad SiO emission (e.g., positions 1, 2, 3, 4, 5, 6, 7, 9, 10, 11, 15, 16, 17, 18, 19, and 20), some of which exhibit red-shifted and/or blue-shifted line wings, and narrow SiO emission (e.g., positions 1, 6, 7, 8, 13, 14, 18, 19, and 20).
Additionally, we note that the SiO emission at positions 2, 9, 10, 11, 12, and 17 is very weak, with peak intensities around the 2$\sigma$ noise levels in Figure~\ref{fig2}. Therefore, we compared these spectra with those from the three circular regions with no detectable SiO emission, as shown in Figure~\ref{figA1}. Except for position~12, the peak intensities of the SiO spectra at the other positions are higher than the 3$\sigma$ noise level measured in the non–emission regions. 
In Table~\ref{tab1}, we summarize the line parameters (peak intensity, $F$; central velocity, ${\upsilon}_{\rm LSR}$; and linewidth, $\Delta {\upsilon}$) for the SiO (2-1) spectra at positions 1 to 20 in Figure~\ref{fig2}. 
These parameters were obtained by fitting the molecular line profiles with Gaussian functions using $\textit{pyspeckit}$. The rms noise level of each spectrum was estimated as the standard deviation over a velocity range of 50 km s$^{-1}$ in line-free channels.
Details of the SiO and H$^{13}$CO$^+$ spectral fitting procedures can be found in Appendix~\ref{AppendixA}.

Figure~\ref{figA2} presents the SiO spectra with their Gaussian fittings, while Figure~\ref{figA3} shows the Gaussian fitting results of the H$^{13}$CO$^+$ spectra at positions 1, 6, 9, 13, and 20, where the spectra do not exhibit absorption features. The H$^{13}$CO$^+$ spectra at the remaining positions display absorption and are therefore shown only as spectra. Upon inspection, we find that the H$^{13}$CO$^+$ spectra show absorption features at positions 2, 3, 4, 5, 7, 8, 10, 11, 12, 14, 15, 16, 17, 18, and 19, which may be due to the missing flux or the presence of strong continuum sources \citep{Ohashi2022ApJ, Codella2024MNRAS}. Notably, positions 3, 4, 5, 7, 8, 18, and 19 are located near outflows (as described in Section~\ref{sec3.1}); positions 9, 10, 11, 12 and 17 fall within the outer arc structure, while positions 13, 14, 15, 16, and 18 lie along the inner arc region. We note that the SiO spectra at the positions in the outer arc are weak and located near the edge of the map, and therefore have larger uncertainties.
The average H$^{13}$CO$^+$ line width at positions 1, 6, 13, and 20, which are located within Filament 2, is 1.6 km s$^{-1}$. In contrast, position 9, situated in the outer arc region, shows the broadest line width of 2.3 km s$^{-1}$. This enhanced linewidth at the outer arc region suggests the presence of strong turbulent motions, possibly driven by CCC or external shocks.

The linewidths of the broad SiO emission range from 3.5 to 28.5 km s$^{-1}$, with an average width of 9.3 km s$^{-1}$, while those of the narrow SiO emission range from 1 to 2.2 km s$^{-1}$, with an average width of 2 km s$^{-1}$. At positions 1, 2, 5, 6, 7, 9, 10, 11, 15, 16, 17, 18, and 19, the SiO linewidths are broader than the broadest H$^{13}$CO$^+$ linewidth (2.3 km s$^{-1}$); however, the line wings are either unresolved or faint, with peak intensities below 2 $\sigma$.  
In contrast, positions 3, 4, and 20 exhibit clear high-velocity wings exceeding 10 km s$^{-1}$ in the red-shifted and/or blue-shifted velocity regimes. These positions are also spatially associated with either the MM cores from \citet{Rathborne2006} or with 8$\,$$\mu$m and 24$\,$$\mu$m sources. Therefore, we propose that the extended high-velocity SiO wings observed in these positions are most likely attributed to outflow activity.  Furthermore, considering the unipolar outflow morphology observed at position 7 and the spatial continuity between the red-shifted lobe at position 20 and the corresponding blue-shifted emission at position 19, we also include positions 7 and 19 as likely outflow candidates.

\begin{figure*}
\centering
\begin{minipage}{1\linewidth}
    \vspace{1pt}
\centering{\includegraphics[width=1\linewidth]{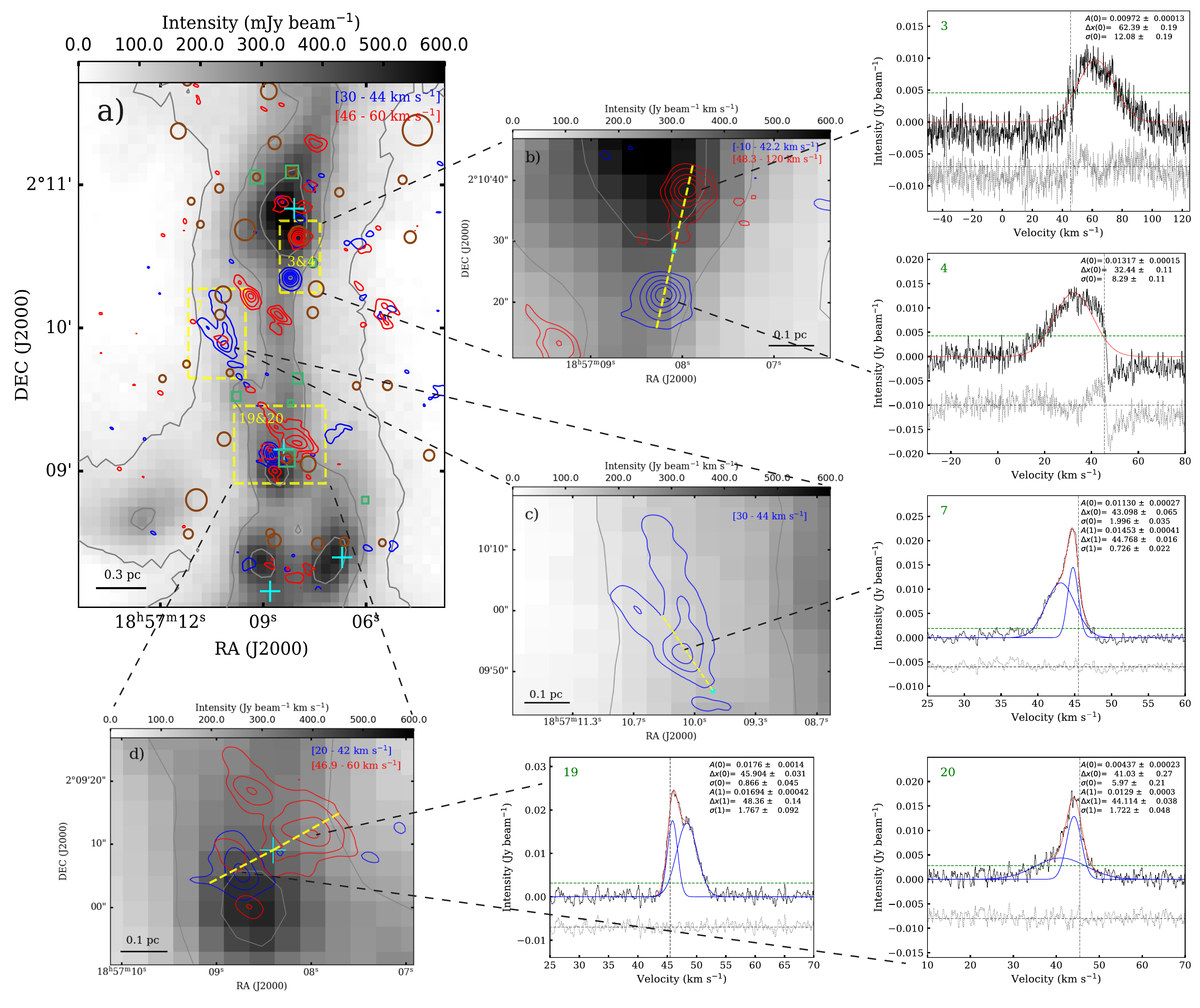}}
 \end{minipage}
\caption{{\it a)}: The background shows the 850 $\upmu$m continuum emission within the ALMA field of view, with contour levels at 55($\sim$5 $\sigma$), 220, 440 mJy beam$^{-1}$. Blue and red contours represent the blue- and red-shifted SiO emission integrated over velocity ranges of 30-44 km s$^{-1}$ and 46-60 km s$^{-1}$, respectively, as indicated in the upper-right corner. The SiO contour levels start at 3$\sigma$ (16 mJy beam$^{-1}$ km s$^{-1}$) and increase by 0.2$I_\mathrm{peak}$ up to $I_\mathrm{peak}$ (910 and 620 mJy beam$^{-1}$ km s$^{-1}$ for the blue and red contours, respectively). 
Yellow dashed rectangles mark the regions corresponding to the green rectangles in Figure~\ref{fig2}, where outflows are identified. The dashed black lines indicate the zoomed-in view of the outflow region. Dark red open circles and medium green rectangles indicate the positions of 8 $\upmu$m sources and 24 $\upmu$m sources, respectively. Other symbols are the same as in Figure~\ref{fig1}. {\it b)}: Zoom-in view of the outflow region at positions 3 and 4. The blue and red velocity ranges are shown in the upper-right corner. Yellow lines indicate the extents of the blue and red lobes, and the cyan star marks the position of the potential driving protostar, located centrally between the lobes. The two right panels indicated by dashed black lines show the SiO spectra with Gaussian fitting extracted from positions 3 and 4. The symbols are the same as Figure~\ref{figA2}. {\it c)}: Zoom-in view of the outflow region at position 7. The velocity range of blue-shifted SiO emission is noted in the upper-right corner. The right panel presents the corresponding SiO spectra with Gaussian fitting. {\it d)}: Zoom-in view of the outflow region at positions 19 and 20. The velocity ranges of blue- and red-shifted SiO emission are shown in the upper-right corner. The two right panels display the SiO spectra with Gaussian fitting extracted from positions 19 and 20.}
\label{fig3}
\end{figure*}

\subsection{Identified outflow}
\label{sec3.1}
In Figure~\ref{fig3} a) panel, we display the blue- and red-shifted SiO emission overlaid with 8 $\upmu$m (dark red open circles) and 24 $\upmu$m (green rectangles) sources.
Around positions 3 and 4, 8~$\upmu$m and 24~$\upmu$m sources are detected $\sim$8$^{\prime\prime}$ to the west, although no continuum core has been identified in our ALMA 3$\,$mm continuum images. At positions 19 and 20, the SiO emission is spatially coincident with both 8~$\upmu$m and 24~$\upmu$m sources and the identified core MM6 \citep{Rathborne2006}.
Based on the observed bipolar morphology of the blue- and red-shifted SiO emission and the associated infrared sources, we confirm that the SiO emission at positions 3, 4, 19, and 20 is due to outflows.
For the outflow at position 7, however, the SiO emission lacks any associated infrared source, although the morphology of this emission resembles a clear outflow cavity. This suggests that the SiO emission toward this position is likely driven by a deeply embedded source. 

\begin{table*}
\renewcommand\tabcolsep{1.1pt}
\centering
\caption{Derived parameters of the SiO (2-1) lines toward the positions selected in IRDC G035.39-00.33.}
\label{tab2}
\begin{threeparttable}
\begin{tabular}{ccccccccccc}
\hline
\hline
Position   & $\int{T_\textup{MB}\mathrm{d}{\upsilon}}$      & $N$(SiO)$\times{10^{12}}$ & $X$(SiO)$\times{10^{-10}}$ & Position   & $\int{T_\textup{MB}\mathrm{d}{\upsilon}}$     & $N$(SiO)$\times{10^{12}}$ & $X$(SiO)$\times{10^{-10}}$ \\
& (mK km s$^{-1}$)    & (cm$^{-2}$) &  &  & 
(mK km s$^{-1}$)       & (cm$^{-2}$) & \\
\hline
1   & 110±4       & 0.19±0.01   & 0.82±0.06  &  
11  & 142±45      & 0.8±0.2     & 1.34±0.33    \\
    & 260±25      & 1.4±0.1     &     
& 
12 & -           & -           &   -           \\
2*  & 153±41      & 0.8±0.2     & 1.8±0.5 &     
13  & 429±22      & 0.74±0.04   & 0.33±0.02    \\
3   & 3738±1023   & 20.3±5.6    & 12.2±3.4     &
14*  & 141±7       & 0.24±0.01   & 0.54±0.02    \\
4   & 3476±541    & 18.9±2.9    & 10.7±1.6     &
15*  & 285±76      & 1.5±0.4     & 3.4±0.9      \\
5   & 1053±192    & 5.7±1       & 10.1±1.8     &
16* & 348±76      & 1.9±0.4     & 4.3±0.9      \\
6   & 296±16      & 0.51±0.03   & 1.52±0.02    &
17* & 451±541     & 2.5±2.9     & 5.7±6.6      \\
    & 439±45      & 2.4±0.2     &             &
18  & 234±89      & 1.3±0.5     & 2.2±0.9      \\
7*   & 718±29      & 3.9±0.2     & 10.2±0.5        &
    & 200±18      & 0.35±0.03   &              \\
    & 336±6       & 0.58±0.01   &              &
19  & 484±14      & 0.84±0.02   & 4.9±0.2      \\
8   & 132±12      & 0.23±0.02   & 0.24±0.02    &
    & 952±40      & 5.2±0.2     &              \\
9   & 339±165     & 1.8±0.9     & 0.9±0.4      &
20  & 830±218     & 4.5±1.2     & 5.6±1.2      \\
10  & 268±80      & 1.5±0.4     & 0.53±0.14    &
    & 704±34      & 1.22±0.06   &  \\
\hline 
\end{tabular}
\end{threeparttable}
 \begin{tablenotes}
    \footnotesize
    \item[] The symbol ‘*’ after a position indicates that the mass surface density at that location is lower than 0.021~g~cm$^{-2}$ (3$\sigma$), for which we adopt 3$\sigma$ as an upper limit.
  \end{tablenotes}
\end{table*}

In Figure~\ref{fig3} panels (b), (c), and (d), we present zoom-in views of the identified outflow regions. The integrated velocity ranges used for the blue- and red-shifted SiO emission are based on their corresponding SiO spectra. 
Panel (b) exhibits intense outflow activity, with high-velocity SiO emission reaching up to terminal velocities of 28 km s$^{-1}$ with respect to the V$_{\rm LSR}$ of the cloud. One infrared source is detected approximately 8$^{\prime\prime}$ to the west, located between the blue and red lobes of the outflow, and another coincides with the red-shifted SiO emission at position 3. 
The outflow in panel (c) presents a possible unipolar morphology with the absence of an infrared source. However, note that its red-shifted counterpart may be overlapping along the line of sight with the red-shifted emission of the molecular outflow found in positions 19 and 20 (see panel d). 
In panel (d), the cyan cross marks the core MM6, which is located between the red and blue lobes, with 8 $\upmu$m and 24 $\upmu$m sources detected. The red-shifted SiO emission presents tails extending toward the northeast, which is likely associated with the red-shifted counterpart of the blue-shifted outflow cavity seen in position 7. The SiO spectra extracted from positions 19 and 20 (see lower-right panel) show two clear blue-shifted and red-shifted velocity components, typical of an outflowing source.
However, given that the physical scale of the ALMA beam at this distance is 0.05 pc, the current resolution is insufficient to resolve the cavity substructures within the SiO emission. Higher-resolution observations will therefore be required to confirm and characterize the presence of multiple outflows.
The extent of the outflow lobes is estimated by measuring the maximum spatial separation between the putative outflow-driving protostars (cyan stars) and the 3$\sigma$ contour levels measured in the SiO emission, as indicated by the yellow lines in the panels of Figure~\ref{fig3}. 
The estimated length of the blue and red lobes ($l_\mathrm{maxb}$ and $l_\mathrm{maxr}$) are $\sim$0.18 pc and 0.2 pc for positions 3 and 4, $\sim$0.18 pc and 0.16 pc for positions 19 and 20, and $\sim$0.21 pc for the blue lobe at position 7. 

By inspecting the outflow properties, we can provide information about the physical properties of the driving sources. The column densities in the outflow regions are estimated as described in Section~\ref{sec3.2}. The SiO column densities ($N_\textup{SiO}$) are 2.0$\times$10$^{13}$ and 1.9$\times$10$^{13}$ cm$^{-2}$ for outflow positions 3 and 4, 4.5$\times$10$^{12}$ cm$^{-2}$ for outflow position 7, and 6.0$\times$10$^{12}$ and 5.7$\times$10$^{12}$ cm$^{-2}$ for outflow positions 19 and 20, respectively.
We adopt a typical SiO abundance ($\chi_\textup{SiO}$) of $\sim$10$^{-8}$, consistent with values reported for mid-infrared-dark sources by \citet{Sanchez-Monge2013A&A}. By converting the $N_\textup{SiO}$ to the H$_2$ column density ($N{_\mathrm{H_{2}}}$), we estimate the masses of the blue and red outflow lobes ($M_\textup{b/r}$), from which the total outflow mass ($M_\textup{out}$) is derived.

\begin{equation}
M_\mathrm{out} = (N_\mathrm{b}\times A_\mathrm{b}+N_\mathrm{r}\times A_\mathrm{r})m_\mathrm{H_2}
\label{eq:eq1}
\end{equation}
\begin{equation}
P = M_\mathrm{b}\lvert V_\mathrm{b}\rvert +M_\mathrm{r} \lvert V_\mathrm{r}\rvert 
\label{eq:eq2}
\end{equation}
\begin{equation}
E = \frac{1}{2}M_\mathrm{b}{V_\mathrm{b}}^2+\frac{1}{2}M_\mathrm{r}{V_\mathrm{r}}^2 
\label{eq:eq3}
\end{equation}
where $A_\mathrm{b/r}$ is the surface area of the blue and red lobes, and $N_\mathrm{b/r}$ is the corresponding $N_\mathrm{H_2}$ column density. $m_\mathrm{H_2}$ represents the mass of a hydrogen molecule. $V_\mathrm{b/r}$ is the velocity offset of the centroid velocities of the blue and red lobes, derived from the Gaussian fitting, with respect to the systemic velocity of Cloud H (45.5 km $^{-1}$). $P$ and $E$ represent the outflow momentum and kinetic energy, respectively. The area $A_\mathrm{b/r}$ is calculated using the same threshold employed for deriving $T_\mathrm{MB}$ values in Table~\ref{tab1}. 
$M_\mathrm{out}$ is estimated to be 1.8 $M_{\odot}$ for positions 3 and 4, and 1.4 $M_{\odot}$ for positions 19 and 20. 
For position 7, we assume that the red lobe has the same $N_\mathrm{r}$ and $A_\mathrm{r}$ as the blue lobe, yielding $M_\mathrm{out}$ = 2.1 $M_{\odot}$. 
$P$ is 27.5 $M_{\odot}$ km s$^{-1}$ for positions 3 and 4, 5.1 $M_{\odot}$ km s$^{-1}$ for position 7, and 5.0 $M_{\odot}$ km s$^{-1}$ for positions 19 and 20. The corresponding $E$ is 4.2$\times{10^{45}}$ erg, 1.2 $\times{10^{44}}$ erg, and 1.9 $\times{10^{44}}$ erg, respectively. These outflow parameters are consistent with those typically found in high-mass protostellar sources \citep{Beuther2002A&A}. 

The parameter $l_\mathrm{max}$ represents the maximum extent of each outflow lobe ($l_\mathrm{maxb/maxr}$), as shown by the yellow lines in Figure~\ref{fig3}. Using this, we estimate the outflow dynamical age ($t_\mathrm{dy}$), mass-outflow rate ($\dot{M}_\mathrm{out}$), mechanical force ($F_\mathrm{SiO}$), and outflow luminosity ($L_\mathrm{SiO}$) using the following equations:
\begin{equation}
\begin{split}
V_\mathrm{maxb/maxr} = V_\mathrm{edge} - V_\mathrm{sys} \\
\end{split}
\label{eq:eq4}
\end{equation}

\begin{equation}
\begin{split}
t_\mathrm{dy} = \frac{l_\mathrm{max}}{(V_\mathrm{maxb}+V_\mathrm{maxb})/2} 
\label{eq:eq5}
\end{split}
\end{equation}
\begin{equation}
\dot{M}_\mathrm{out} = \frac{M_\mathrm{out}}{t}
\label{eq:eq6}
\end{equation}
\begin{equation}
F_\mathrm{SiO} = \frac{P}{t}
\label{eq:eq7}
\end{equation}
\begin{equation}
L_\mathrm{SiO} = \frac{E}{t}
\label{eq:eq8}
\end{equation}

where V$_\mathrm{edge}$ corresponds to the most blue-shifted or red-shifted velocity, and V$_\mathrm{sys}$ is the systemic velocity of Cloud H (45.5 km$^{-1}$). 
The maximum velocity offsets, $V_\mathrm{maxb/r}$, are defined as the velocity offset between $V_\mathrm{edge}$ and $V_\mathrm{sys}$. 
The resulting values are: V$_\mathrm{maxb}$ = 32.6 km s$^{-1}$ and V$_\mathrm{maxr}$ = 45.4 km s$^{-1}$ for positions 3 and 4; V$_\mathrm{maxb}$ = 7.1 km s$^{-1}$ for position 7; and V$_\mathrm{maxb}$ = 18.6 km s$^{-1}$ and V$_\mathrm{maxr}$ = 7.1 km s$^{-1}$ for positions 19 and 20.
To correct for the unknown inclination angle between the outflow axis and the line of sight, we assume an average inclination angle of $\theta$ = 57.3$^{\circ}$ \citep{Beuther2002A&A}. 
The estimated $l_\mathrm{max}$ values are $\sim$0.37 pc for positions 3 and 4, $\sim$0.39 pc for position 7, and $\sim$0.33 pc for positions 19 and 20.
The corresponding $t_\mathrm{dy}$ are estimated to be 9.3$\times$${10^{3}}$, 5.4$\times{10^{4}}$, and 2.5$\times{10^{4}}$ yr, respectively. \citet{Liu2021ApJ} identified outflows using SiO (5–4) emission toward Cloud H, and their calculated $t_\mathrm{dy}$ is approximately 5$\pm$4 $\times$10$^3$ yr, which is younger than the outflows we have identified.
The $\dot{M}_\mathrm{out}$ are 2$\times{10^{-4}}$, 3.9$\times{10^{-5}}$, and 5.4$\times{10^{-5}}$ $M_{\odot}$ yr$^{-1}$, respectively. 
The $F_\mathrm{SiO}$ are derived to be 3$\times{10^{-3}}$, 9.4$\times{10^{-5}}$, and 2$\times{10^{-4}}$ $M_{\odot}$ km s$^{-1}$ yr$^{-1}$ for the same positions.
The derived $L_\mathrm{SiO}$ for positions 3, 4, 7, 19, and 20 are 3.7 $L_\odot$, 0.02 $L_\odot$, and 0.06 $L_{\odot}$, respectively. We adopt a ratio between the mass loss rate of the jet and the accretion rate of 0.3, following \citet{Tomisaka1998ApJ, Shu1999ASIC, Beuther2002A&A}, and drive the accretion rate with the spread between 1.3$\times{10^{-4}}$ $M_{\odot}$ yr$^{-1}$ and 6.5$\times{10^{-4}}$ $M_{\odot}$ yr$^{-1}$, which are consistent with typical values observed in high-mass protostellar cores \citep{Beuther2002A&A, Yang2018ApJS}. 

\begin{figure*}
\begin{minipage}{1\linewidth}
    \vspace{3pt}
\centerline{\includegraphics[width=1.1\linewidth]{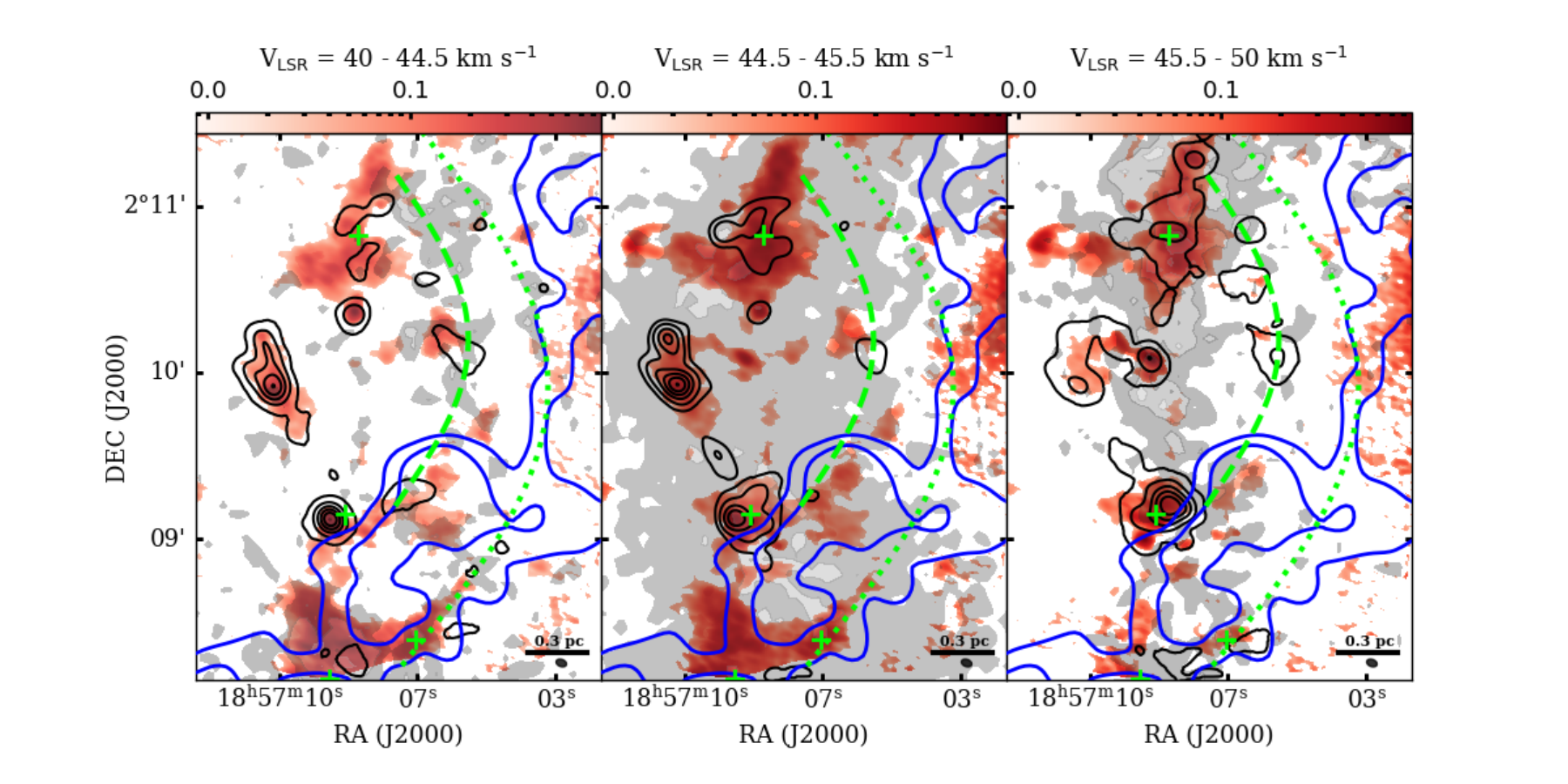}}
 \end{minipage}
\caption{The red background represents the CS integrated intensity, overlaid with SiO (2-1) emission (black contours) and C$^{18}$O (2-1) emission (gray contours). The integrated velocity ranges from 40 to 44.5 km s$^{-1}$ (blue-shifted gas; left panel), 44.5 to 45.5 km s$^{-1}$ (ambient gas; central panel), and 45.5 to 50 km s$^{-1}$ (red-shifted gas; right panel).
SiO (2–1) black contour levels start at 3$\sigma$ and increase by 0.2$I_\mathrm{peak}$ up to $I_\mathrm{peak}$. The rms noise levels ($\sigma$) for the different velocity components are 7.4, 7.5, and 7.4~Jy~beam$^{-1}$~km~s$^{-1}$, and the corresponding peak intensities ($I_\mathrm{peak}$) are 1010, 288, and 877~Jy~beam$^{-1}$~km~s$^{-1}$, respectively. The C$^{18}$O gray contour levels in the left panel are 0.738, 1.107, 1.476, and 1.738 K km s$^{-1}$ (peak value). In the central and right panels, the contour levels are 0.322, 0.805, 1.127, and 1.44 K km s$^{-1}$ (peak value), and 0.624, 0.936, 1.248, 1.56, and 1.75 K km s$^{-1}$ (peak value), respectively. The blue contours display 610~$\mathrm{MHz}$ emission at 1.1 (5$\sigma$) and 1.54 (7$\sigma$) mJy beam$^{-1}$ from \citet{Paredes2014A&A}. Green crosses mark the locations of the identified cores. Green dashed and dotted lines indicated the arc-like structures. The beam size and physical scale are displayed in each lower right corner.}
\label{fig4}
\end{figure*}  

\subsection{Column density and abundance}
\label{sec3.2}
We estimate SiO column density ($N_\textup{SiO}$) and abundance with respect to H$_2$ toward all positions. Assuming local thermodynamic equilibrium (LTE) and that the SiO emission is optically thin \citep{jimenez2010parsec}, the column density of SiO can be calculated as follows: 
\begin{equation}
N = \frac{3 k^2}{4 \uppi^3 h \nu^2} \frac{1}{S{\mu}^2}{T_\textup{ex}}
{{\rm e}^{\frac{E_{u}}{kT_\textup{ex}}}}{\int{T_\textup{MB}\mathrm{d}{\upsilon}}}
\label{eq:eq9}
\end{equation}

where $h$ represents the Planck constant and $k$ is the Boltzmann constant. The line strength ($S\mu^{2}$) is 19.2 for the SiO (2-1) line, and the SiO line frequency ($\nu$) is 86.847 GHz. The upper-level energy $E_u/k$ is 6.25 K. 
We utilized an excitation temperature ($T_\textup{ex}$) of 9 K for narrow components, as estimated for Cloud H \citep{jimenez2010parsec}, and 50 K for the broad component, as estimated for the shocked gas in molecular outflows \citep{2005Jimnez}.
The term $\int{T_\mathrm{MB}\mathrm{d}{\upsilon}}$ represents the integrated intensity of SiO emission in units of [mK km s$^{-1}$]. The values of $\int{T_\textup{MB}\mathrm{d}{\upsilon}}$ and $N_\textup{SiO}$ are provided in Table~\ref{tab2}.

\begin{figure}
\centering
\begin{minipage}{1.1\linewidth}
    \vspace{3pt}
\centerline{\includegraphics[width=0.95\linewidth]{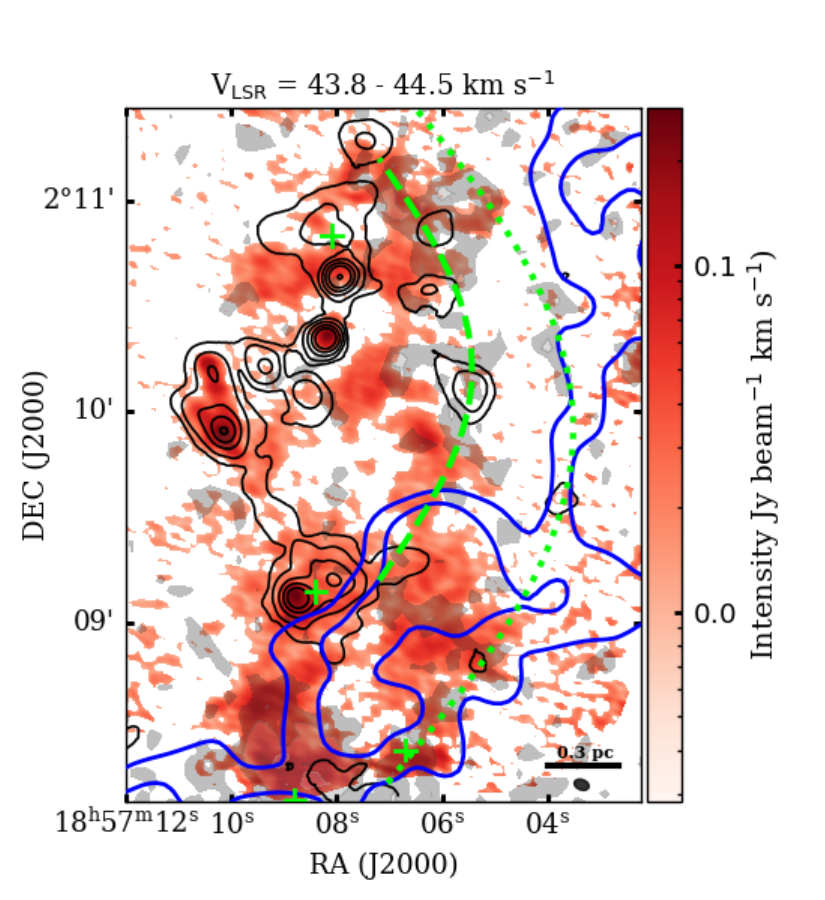}}
 \end{minipage}
\caption{The red background represents the CH$_3$OH integrated intensity, overlaid with SiO (2-1) emission (black contours) and C$^{18}$O (1-0) emission (gray contours). The integrated velocity ranges from 43.8 to 44.5 km s$^{-1}$, which represents the blue-shifted gas.
SiO black contour levels are 21 (3$\sigma$), 210, 630, 1050, and 2300 (peak value) mJy beam$^{-1}$ km s$^{-1}$. The C$^{18}$O gray contour levels are 0.738, 1.107, 1.476, and 1.738 K km s$^{-1}$ (peak value). The blue contours display 610~$\mathrm{MHz}$ emission at 1.1 (5$\sigma$) and 1.54 (7$\sigma$) mJy beam$^{-1}$ from \citet{Paredes2014A&A}. Green crosses mark the locations of the identified cores. Green dashed and dotted lines indicated the arc-like structures. The beam size is displayed in the lower right corner.}
\label{fig5}
\end{figure} 

For the identified outflow regions (positions 3, 4, 7, 19, and 20), $N_\textup{SiO}$ ranges from 4.5$\times{10^{12}}$ to 2.0$\times{10^{13}}$ cm$^{-2}$. In the adjacent outflow regions (positions 1, 5, 6, and 8), $N_\textup{SiO}$ ranges from 2.3$\times{10^{11}}$ to 5.7$\times{10^{12}}$ cm$^{-2}$. The average $N_\textup{SiO}$ in the outflow and adjacent regions is 7.3$\times{10^{12}}$ cm$^{-2}$.
For the SiO inner arc-shaped regions (positions 13, 14, 15, 16, and 18), $N_\textup{SiO}$ ranges from 2.4$\times{10^{11}}$ to 1.9$\times{10^{12}}$ cm$^{-2}$, with an average $N_\textup{SiO}$ of 1.2$\times{10^{12}}$ cm$^{-2}$.
For the SiO outer arc-shaped regions (positions 9, 10, 11, and 17), $N_\textup{SiO}$ ranges from 8$\times{10^{11}}$ to 2.5$\times{10^{12}}$ cm$^{-2}$, with an average $N_\textup{SiO}$ of 1.7$\times{10^{12}}$ cm$^{-2}$. Note that the derived column densities of the narrow SiO change by less than a factor of 1.3 when assuming an excitation temperature of 15 K instead of 9 K, and those of the broad SiO column vary by more than a factor of 1.4 when assuming 75 K instead of 50 K.

To estimate the SiO abundance ($\chi_\textup{SiO}$), we calculate the H$_2$ column density using the mass surface density at the same 20 positions. For the position (2, 7, 14, 15, 16, and 17) with the mass surface density lower than 0.021~g~cm$^{-2}$ (3$\sigma$), we adopt 3$\sigma$ as an upper limit. The derived $\chi_\textup{SiO}$ values are presented in Table~\ref{tab2} and range from 2.4$\times10^{-11}$ to 1.2$\times10^{-9}$. 
For the identified outflow and adjacent regions, the average $\chi_\textup{SiO}$ is $\sim$6.3$\times{10^{-10}}$. The average $\chi_\textup{SiO}$ in the inner arc-shaped regions is $\sim$2.2$\times{10^{-10}}$, while in the outer arc-shaped regions, it is $\sim$2.1$\times{10^{-10}}$. These results imply that the SiO abundances in both the inner and outer arcs are lower by a factor of $\sim$3 compared to those measured in the outflow regions. As discussed in Section \ref{sec4}, this may indicate a different shock origin for the inner and outer arcs. However, note that these values should be considered as lower limits since the SiO images suffer from missing flux (see Section \ref{sec2}). 
                        
\begin{figure*}
\centering
\begin{minipage}{1\linewidth}
    \vspace{3pt}
\centerline{\includegraphics[width=1.5\linewidth]{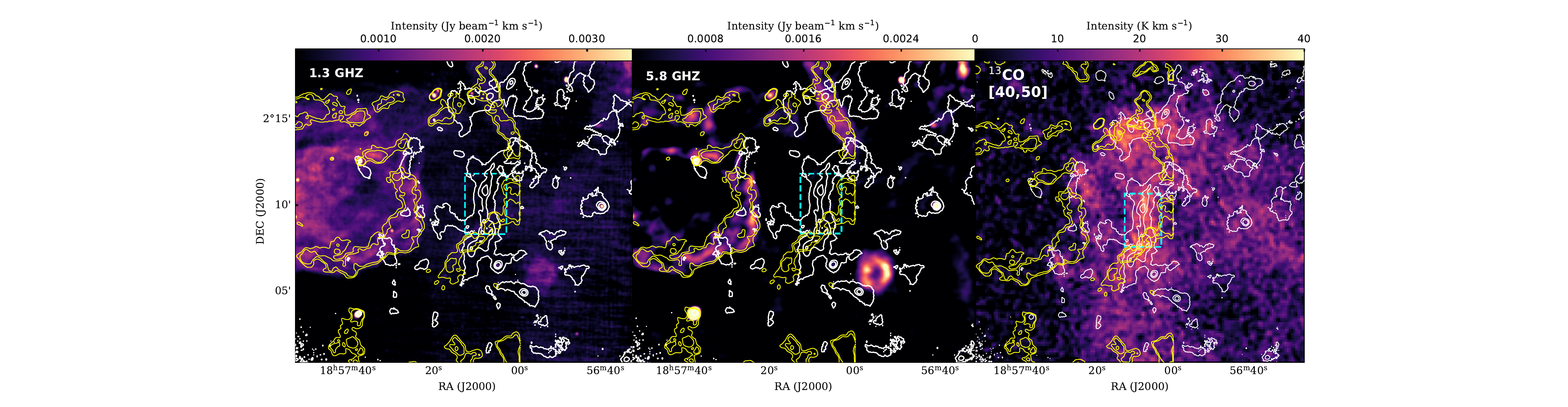}}
 \end{minipage}
\caption{{\it Left panel}: The background displays 1.3~$\mathrm{GHz}$ emission obtained from \citet{Goedhart2024MNRAS}, overlaid with yellow contours representing 610~$\mathrm{MHz}$ emission at levels of 1.1 (5$\sigma$) and 1.54 (7$\sigma$) mJy beam$^{-1}$ from \citet{Paredes2014A&A}. White contours represent the 850 $\upmu$m continuum emission obtained with JCMT/SCUBA-2 \citep{Shen2024}, with levels at 55, 220, 440 mJy beam$^{-1}$. The cyan rectangle indicates the field of view for the ALMA observation. {\it Middle panel}: The background displays 5.8~$\mathrm{GHz}$ emission obtained from \citep{Medina2019A&A}, with yellow contours representing 610~$\mathrm{MHz}$ emission. Other symbols are the same as those in the left panel. {\it Right panel}: The background shows $^{13}$CO (1-0) emission from the FUGIN survey \citep{Umemoto2017PASJ}, with yellow contours representing 610~$\mathrm{MHz}$ emission. The integrated velocity range is shown in the upper-left corner. Other symbols are the same as those in the left panel.}
\label{fig6}
\end{figure*}

\section{Discussion}
\label{sec4} 
The formation of molecular clouds, particularly the filamentary structures found in IRDCs, has long been a topic of interest to understand the early stages of massive star and cluster formation.
Several theoretical models have been proposed to explain their formation. Among these, the flow-driven formation models suggest that flows of warm atomic gas may converge at relatively high velocities and molecular material may form as a result of gas cooling at the converging layer \citep{benjamin2003,vanLoo2007,Hennebelle2008A&A}. However, this model faces significant challenges, as magnetic fields can inhibit compression to the high densities required for star formation \citep{Kortgen2015MNRAS}. Another possible mechanism is shock-induced star formation triggered by stellar feedback \citep{Koyama2000, Koyama2002, vanLoo2007, Inutsuka2015A&A}. In this scenario, molecular clouds and/or clump already formed may be further compressed by shocks driven by the expanding shells of H\textsc{ii} regions and SNRs, setting favourable conditions for star formation to occur \citep{Inoue2018PASJ}. In this shock-driven scenario, narrow SiO emission can be witnessed \citep{Cosentino2019}, with features similar to those expected in cloud-cloud collisions \citep{jimenez2010parsec, 2018Cosentino, Cosentino2020}.

Previous studies point to a cloud-cloud collision as the likely mechanism responsible for the formation of the IRDC G035.39-00.33.
\citet{jimenez2010parsec} indeed found widespread narrow SiO emission across parsec-scales toward this filament, which was interpreted as a fossil record of the cloud-cloud collision within Cloud H. By using $^{13}$CO (1–0) and C{\small II} emission, \citet{Liu2018ApJ, Bisbas2018MNRAS} proposed that a large-scale cloud-cloud collision could be causing the observed velocity gradients and the collision of the filaments. However, the origin of such a collision remains unknown. By carrying out a detailed analysis of the kinematics of the molecular emission of $^{13}$CO and C$^{18}$O, \citet{jimenez2014} found three velocity-coherent molecular filaments within the cloud with a common velocity gradient of $\sim$0.4-0.8 km s$^{-1}$ pc$^{-1}$ in the north-south direction. These authors proposed that the common velocity gradient could be explained by the global gravitational collapse of the cloud, which would induce the gentle collision of the individual filaments, yielding the observed widespread and narrow SiO emission \citep{jimenez2014}. These observations, however, lacked high enough angular resolution. Additionally, the presence of velocity-coherent fibres in the IRDC may also support a shock-driven scenario. 
Numerical simulations by \citet{Clarke2017MNRAS,Clarke2020MNRAS} show that fibres can arise from internal turbulence generated by shocks, which convert inflowing kinetic energy into turbulent motions within filaments. The observed fibres could hence be kinematic artefacts rather than true density structures \citep{Clarke2018MNRAS}.

In Figure~\ref{fig2}, we report an arc-like structure in SiO emission located outside the densest material of cloud H as seen in extinction. 
In Figure~\ref{fig4}, we present the CS integrated intensity maps overlaid on the observed SiO emission for different velocity ranges. The brightest CS emission is detected for the central component between 45.5 and 46.5 km s$^{-1}$. However, for the blue-shifted (44.5 to 45.5 km s$^{-1}$) and red-shifted (46.5 to 48 km s$^{-1}$) CS gas components, one can see two arc-like structures (indicated with green dashed and dotted lines) extending from south to north that closely follow the SiO emission. 
While one arc is located on the inner side (inner arc), the other one is positioned along the edge of the map (outer arc). 
To confirm that these structures are real, in Figure~\ref{fig5} we also present the integrated intensity map of another shock tracer, CH$_3$OH, over the blue-shifted velocity range 43.8-44.5 km s$^{-1}$. The red color scale highlights both the inner arc and outer arcs, as indicated by the green dashed and dotted lines.
Additionally, we include the C$^{18}$O emission observed with the GBT telescope in Figures~\ref{fig4} and \ref{fig5} (Law et al. in prep.), shown in grayscale. The C$^{18}$O data have a beam size of 7.7$^{\prime\prime}$, a velocity resolution of 0.19~km~s$^{-1}$, and an rms noise level of 0.47~K per channel. The blue-shifted C$^{18}$O emission, represented by Filament 1 (see Section~\ref{sec1}), also follows these arc-like structures. 
In the next sections, we discuss possible scenarios that could explain the arc-like structures seen in SiO, CH$_3$OH, and CS emission with ALMA toward Cloud H. 

\subsection{Interaction with Supernova Remnants}
\citet{Cosentino2019} analyzed ALMA images of SiO (2-1) emission toward the SNR W44. The authors reported evidence of non-stationary MHD shocks driven by the SNR into the nearby IRDC G034.77-00.55. Interestingly, Cloud H is also located nearby the SNR G35.6-0.6 \citep{Paredes2014A&A}. Initially reported to be at a kinematic distance of 3.6 kpc, subsequent CO observations \citep{Zhang2022ApJ} have shown that the SNR is composed of two shells located at 3.4 and 3.0 kpc, respectively. The shell at 3 kpc is not only at the same distance as Cloud H but it is also located closest on the plane of the sky. Therefore, in this section, we explore the feasibility of the arc-like structures being produced by the interaction between Cloud H and the nearby SNR G35.6-0.6 \citep{Paredes2014A&A}. 
 
Figure~\ref{fig6} shows the MeerKAT 1.3 $\mathrm{GHz}$ image \citep{Goedhart2024MNRAS}, the GLOSTAR 5.8 $\mathrm{GHz}$ image \citep{Medina2019A&A}, and the FUGIN $^{13}$CO (1-0) image \citep{Umemoto2017PASJ}. The images are overlaid with the 610 $\mathrm{MHz}$ emission (shown as yellow dashed contours) obtained with the GMRT by \citet{Paredes2014A&A}.
These radio frequencies trace both non-thermal synchrotron emission, particularly at lower frequencies such as 610~MHz, and thermal free-free emission from ionized gas, which becomes more prominent at higher frequencies such as 5.8~GHz. 
The white contours represent the 850 $\upmu$m continuum emission obtained with JCMT/SCUBA-2 \citep{Shen2024}. The overall appearance of the 1.3~$\mathrm{GHz}$ emission closely resembles that of the 610~$\mathrm{MHz}$ emission, while the 5.8~$\mathrm{GHz}$ emission shows similar but comparatively weaker structures. Owing to the higher angular resolution of the 1.3~$\mathrm{GHz}$ and 5.8~$\mathrm{GHz}$ observations, more detailed features are revealed. 
The bright rings associated with SNR G35.6-0.4 are visible at both wavelengths as shown in Figure~\ref{fig6}, and are located to the east of Cloud H. Note that the 610~$\mathrm{MHz}$ emission lacks coverage westward of $\alpha$(J2000)=18:57:00. Nevertheless, based on the available data not only at 610 MHz but also at 1.3~$\mathrm{GHz}$ and 5.8~$\mathrm{GHz}$, we find two possible shell structures related to SNR G35.6-0.4, one of which spatially overlaps with Cloud H not only in the plane of the sky but it is also located at the same distance of 3.0 kpc \citep{Zhang2022ApJ}. Interestingly, the arc-shaped structure seen in SiO, CH$_3$OH, and CS in the ALMA images of cloud H has the same spatial orientation as the morphology of the SNR shell (see Figures \ref{fig4} and \ref{fig5} and discussion below).
At 1.3 GHz and 5.8 GHz maps, we also observe a second spherical shell located toward the south-western part of Cloud H. This object, located at $\alpha$(J2000)=18:56:55, $\delta$(J2000)=02:06:03, exhibits a brighter, circular morphology in 5.8 GHz emission compared to other radio emission, suggesting that it is likely an H\textsc{ii} region.

In the right panel of Figure \ref{fig6}, the $^{13}$CO (1-0) emission closely traces the dust continuum emission (shown as white contours), revealing the overall morphology of the giant molecular clouds in this region. These observations indicate that Cloud H is embedded within a complex star-forming environment, where the densest and coolest material appears to be concentrated at the interface between SNR G35.6-0.4 and the expanding H\textsc{ii} region. Note that the $^{13}$CO (1-0) map has been obtained for the velocity range between 40 and 50 km s$^{-1}$, which is also the velocity range of the molecular gas in Cloud H.

If we zoom-into the second shell of SNR G35.6-0.4, we observe that the 610~$\mathrm{MHz}$ emission (blue contours in Figures~\ref{fig4} and \ref{fig5}, and yellow contours in the left and middle panels of Figure~\ref{fig6}) delineates the morphology of Filament 1 in Cloud H, as reported by \citet{jimenez2014} and traced by $^{18}$CO (1–0) emission (see the left panel grayscale in Figure~\ref{fig4} and \ref{fig5}). Note that the dense gas of Cloud H bends toward the southern part of the cloud following the morphology of the second shell of SNR G35.6-0.4, suggesting a possible interaction. Furthermore, the observed SiO, CS, and CH$_3$OH arc structures follow the outline of the 610~$\mathrm{MHz}$ emission, which may further support the interaction between the SNR G35.6-0.4 and Cloud H. A possible explanation is that the expansion of SNR G35.6-0.4 produced a shock interaction with the preexisting Cloud H. This interaction shaped the arc structure of Filament 1 seen in SiO, CH$_3$OH, and CS in the same fashion as the outer shell of SNR G35.6-0.4. The fact that the associated SiO abundances are factors of at least 3 lower than those detected in outflows, supports a scenario of a gentle interaction with the expanding swept-up material from the SNR. 
However, due to the limited westward coverage of the 610~$\mathrm{MHz}$ observations, the actual overlap between the radio emission from SNR G35.6–0.4 and Cloud H remains uncertain. Furthermore, it is unclear whether the mass and momentum of the swept-up gas are sufficient compared to those of the IRDC and whether this scenario can account for the observed deuteration \citep{Barnes2016} and the relatively quiescent kinematics in the filament \citep{jimenez2014, Henshaw2013}. Therefore, at this stage it is not possible to undoubtedly confirm this scenario.

We finally note that the overlapping radio emission could originate from the second spherical shell located toward the south-western region of Cloud H. This object, which could be either another SNR or an H\textsc{ii} region, could also be interacting with Cloud H. We have measured the integrated flux across this source at 1.3 GHz \citep[obtained with MeerKAT;][]{Goedhart2024MNRAS} and at 5.8 GHz \citep[GLOSTAR;][]{Medina2019A&A}, deriving 1.4 mJy and 7 mJy, respectively. The inferred spectral index is $\alpha$$\sim$1, which suggests that this spherical shell is likely associated with an optically thick H\textsc{ii} region. 
This object has indeed been classified as an H\textsc{ii} region \citep[see Figure 1 in][]{Shen2024}. 
Furthermore, radio recombination lines from this object show a velocity similar to that of Cloud H, indicating that they belong to the same star-forming cloud complex. Therefore, an interaction between the expanding gas of the H\textsc{ii} region and Cloud H could be possible. Future observations of shock tracers, such as SiO, toward the southern part of Cloud H will be crucial to assess whether such large-scale interaction is currently ongoing \citep{Cosentino2025A&A}. 

\subsection{Global gravitational collapse and cloud-cloud collisions}
As mentioned above, an alternative scenario proposed by \citet{jimenez2014} is that the molecular filaments in Cloud H are undergoing gravitational collapse and as they approach to each other, they experience a cloud-cloud collision. This collision, which would be more gentle than in the SNR scenario, would be responsible for the widespread and narrow SiO emission. 
\citet{Hernandez2015ApJ} used $^{13}$CO observations to find disturbed kinematics in Cloud H, suggesting that GMC–GMC collisions can trigger star formation within the cloud. Similarly, \citet{Liu2018ApJ} proposed that Cloud H may have formed through a large-scale ($\sim$10 pc) cloud–cloud collision, as indicated by the velocity gradients observed in $^{13}$CO emission.
Furthermore, \citet{Bisbas2018MNRAS} supported this scenario by comparing their modeled [C$_\mathrm{II}$] emission with observations, finding results consistent with predictions from the colliding-cloud scenario.

In our results, the shock tracers exhibit spatially extended emission with relatively narrow linewidths, indicating that they may originate from low-velocity shocks. Previous studies have also reported quiescent kinematics within Cloud H \citep{Henshaw2013}, further suggesting the presence of gentle dynamical processes involved in its formation. Taken together, these findings provide an alternative scenario in which both the observed arc-like structures and the formation of Cloud H may be the result of a collision between giant molecular clouds. However, note that this scenario does not explain the morphological coincidence between the 610 MHz emission associated with the SNR G35.6-0.4 and the arc-like structures seen with ALMA in SiO, CH$_3$OH, and CS, as described in Section 4.1.

\subsection{A widespread population of low-mass protostars?}
A final scenario that could explain the observed narrow SiO in Cloud H is the presence of a population of low-mass protostars. Narrow SiO would then arise from either decelerated shocked gas from large-scale outflows, as proposed by \citet{lefloch1998widespread, Beuther2007ApJ}, or from material recently processed in the magnetic precursor of MHD shocks from young molecular outflows \citep{2004Jimnez,2005Jimnez, Rong2025}. However, note that the arc-like structures are found away from the densest material in Cloud H, and therefore, the probability of finding such a low-mass protostellar population outside the main filament of Cloud H seems rather unlikely. In addition, as shown in Figure \ref{fig1}, there are not 8 $\mu$m sources associated with the arc-like structures, which supports the idea of these structures not being produced by protostellar activity.

Future ALMA observations, including 12m-array, ACA, and TP observations with higher sensitivity, are needed to determine which mechanism of the ones presented above is primarily responsible for the observed SiO arc-like structures detected toward Cloud H.
\section{Conclusions}
\label{sec5}
We have conducted an analysis of SiO emission toward the filamentary IRDC G035.39-00.33 using ALMA data, complemented by observations of C$^{18}$O, CH$_3$OH, and CS to trace shock signatures within this cloud. Our main conclusions are as follows:
\begin{enumerate}
      \item {From ALMA SiO observations, we have identified three outflows within the cloud. The derived outflow properties suggest that we are witnessing the formation of a cluster of high-mass stars, which provides clear evidence for ongoing star formation in these cores.} 
      \item {The presence of two large-scale arc-like structures seen in SiO, CH$_3$OH, and CS emission in the high-angular resolution ALMA images, along with C$^{18}$O blue-shifted gas, provides strong evidence of a large-scale shock interaction within the filamentary G035.39-00.33 cloud. These shocks may have been induced by larger-scale processes.}
      \item {Using MeerKAT 1.3 $\mathrm{GHz}$, GLOSTAR 5.8 $\mathrm{GHz}$, FUGIN $^{13}$CO emission, and 610 $\mathrm{MHz}$ GMRT observations, we suggest that the expansion of SNRs and cloud-cloud collisions may be responsible for the arc-like structures observed in SiO, CH$_3$OH, and CS emission. These structures indicate that external shocks have compressed the molecular gas, shaping and compressing the dense ridge of Cloud H.}
\end{enumerate}

\begin{acknowledgements}
We would like to thank Prof. J. M. Paredes for kindly sharing his GMRT 610 MHz image of the SNR G35.6-0.4. R.L., G.C. and I.J.-S. acknowledge funding from grant No. PID2022-136814NB-I00 from MICIU/AEI/10.13039/501100011033 and by “ERDF—A way of making Europe”. This publication makes use of data from FUGIN, the FOREST Unbiased Galactic plane Imaging survey with the Nobeyama 45-m telescope, a legacy project in the Nobeyama 45-m radio telescope. T.L. acknowledges the supports by the National Key R\&D Program of China (no. 2022YFA1603100), National Natural Science Foundation of China (NSFC) through grants no. 12073061 and no. 12122307, and the Tianchi Talent Program of Xinjiang Uygur Autonomous Region.
\end{acknowledgements}

\bibliographystyle{aa}   
\bibliography{paper}
\appendix{}
\section{Details of the spectral fitting}
\label{AppendixA}
\begin{figure*}
\centering
 \includegraphics[width=1\textwidth]{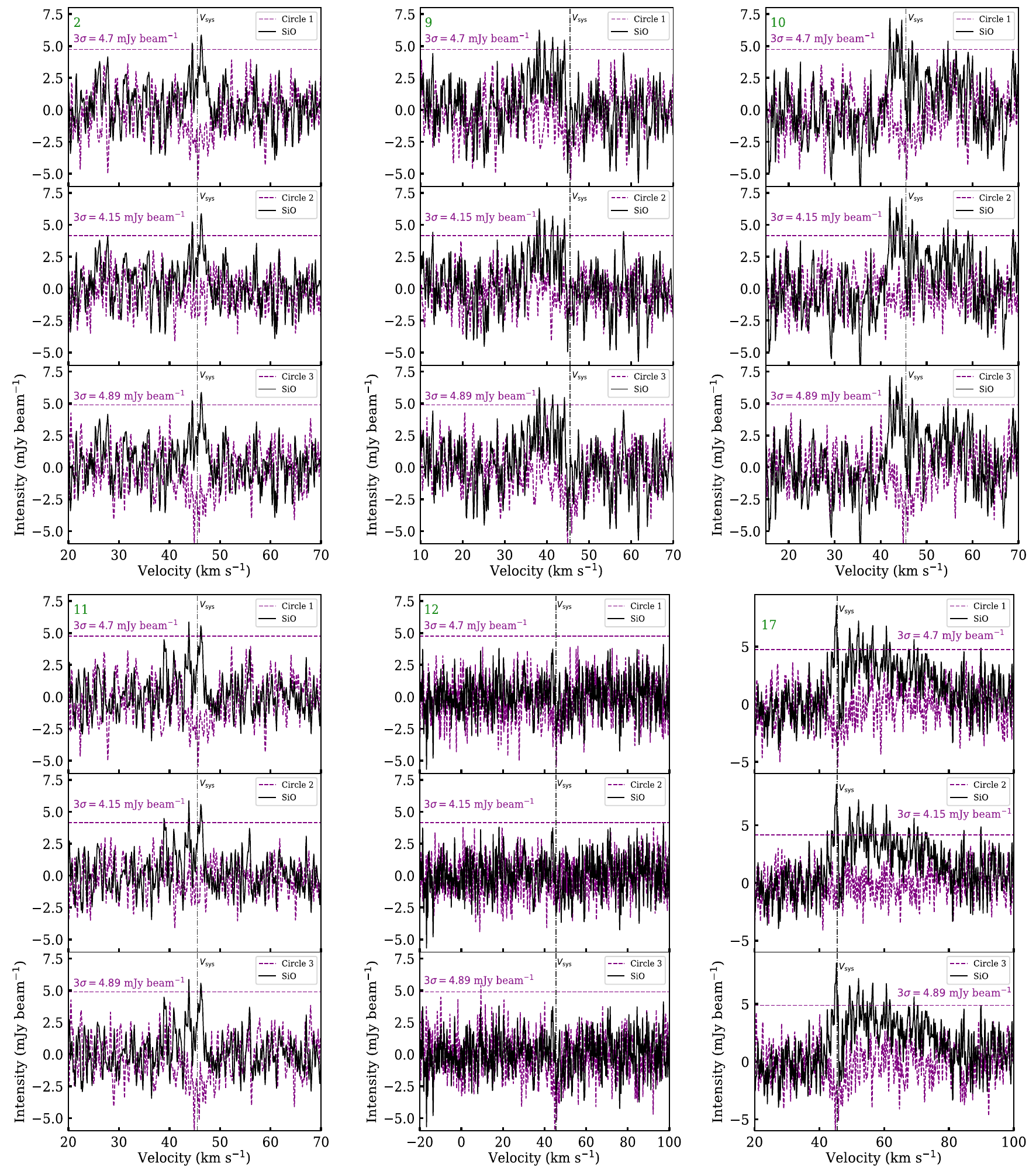}
  \caption{Black curves represent the SiO spectra extracted from the positions 2, 9, 10, 11, 12, and 17. The purple curves show the SiO spectra extracted from the purple circular regions with no SiO emission, as indicated in Figure~\ref{fig2}. The numbers at the upper right corner of each panel correspond to the labels of these circular regions. The purple dashed parallel lines indicate the 3$\sigma$ noise levels of the SiO emission from non-emission regions. The vertical dashed-dotted black line marks the central velocity of Cloud H~(45.5 km $^{-1}$).}
  \label{figA1}
\end{figure*}

Figure~\ref{figA1} presents the low signal-to-noise SiO spectra extracted from positions 2, 9, 10, 11, 12, and 17, compared with those from the non–SiO emission regions (purple circles 1, 2, and 3, as indicated in Figure~\ref{fig2}), respectively.
Figure~\ref{figA2} shows the Gaussian fitting of the molecular line emission using $\textit{pyspeckit}$ for positions 1–20, except position 12, where no emission is detected. Figure~\ref{figA3} presents the Gaussian fitting results of the H$^{13}$CO$^+$ spectra at positions 1, 6, 9, 13, and 20, where the spectra do not show absorption features. The H$^{13}$CO$^+$ spectra at the other positions exhibit absorption and are therefore shown only as spectra.

We obtained the central velocity (${\upsilon}_{\rm LSR}$, km s$^{-1}$), linewidth ($\Delta {\upsilon}$ km s$^{-1}$), and peak intensity ($F$ mJy beam$^{-1}$) from $\textit{pyspeckit}$.
These parameters, including the uncertainties and the noise levels, are listed in Table~\ref{tab1}. The rms noise levels of each spectrum were estimated from the standard deviation over a velocity range of 50 km s$^{-1}$ in line-free channels. The corresponding uncertainties ($\sigma_{\rm I}$, mJy beam$^{-1}$) were derived as follows:
\begin{equation}
\sigma_{\rm I} = \sigma_{\rm rms}\times\sqrt{N_{\rm chan}}\times\Delta {\upsilon}. 
\label{eqA1}
\end{equation}
where $\sigma_{\rm rms}$, $N_{\rm chan}$, and $\Delta {\upsilon}$ represent the rms noise level, the number of channels used in the Gaussian fitting, and the velocity width per channel, respectively. 

\begin{figure*}
\centering
\begin{minipage}{0.33\linewidth}
    \includegraphics[width=\linewidth]{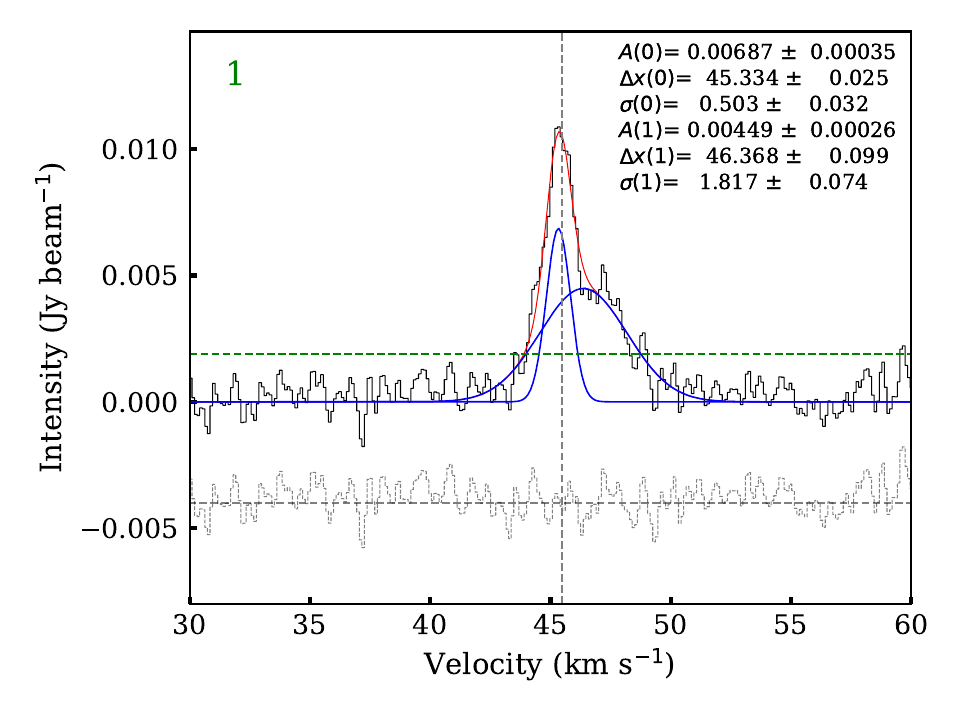}
\end{minipage}
\begin{minipage}{0.33\linewidth}
    \includegraphics[width=\linewidth]{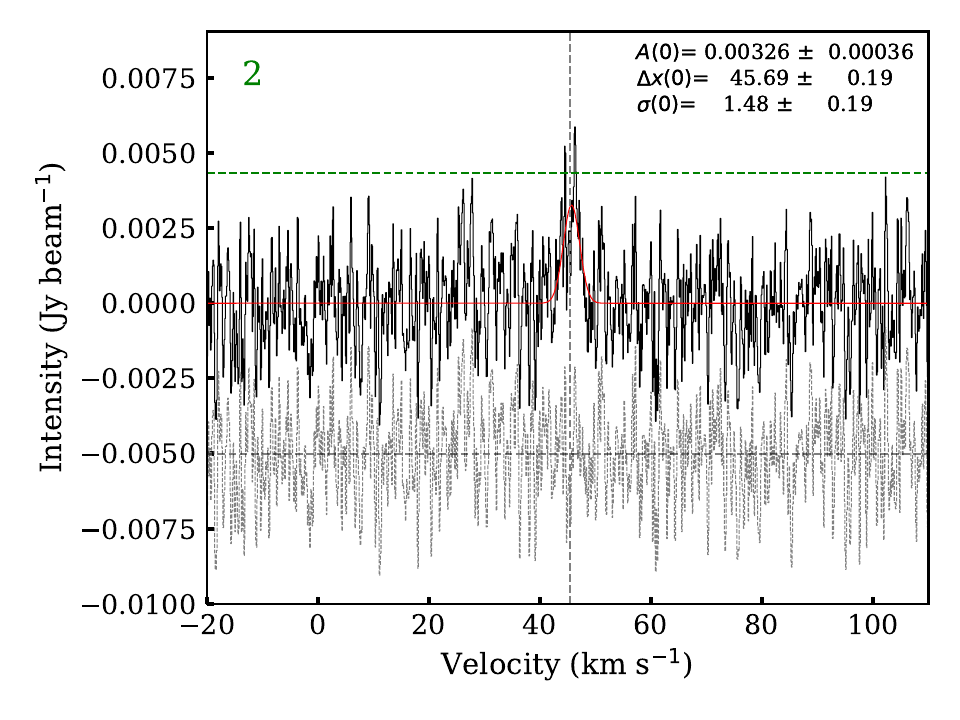}
\end{minipage}
\begin{minipage}{0.33\linewidth}
    \includegraphics[width=\linewidth]{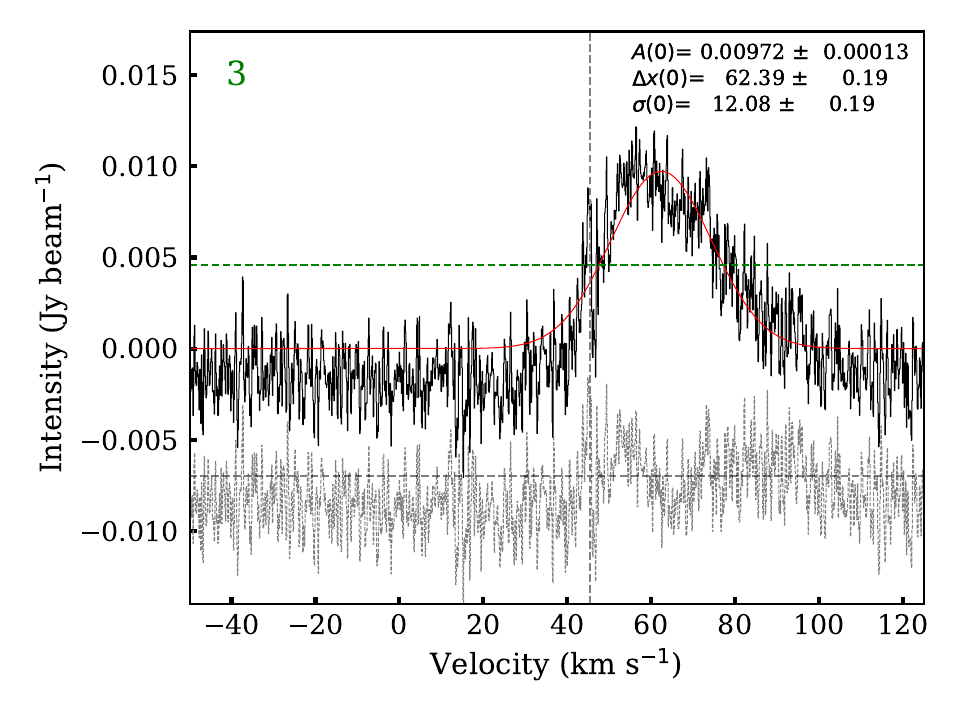}
\end{minipage}
\vspace{5pt} 
\begin{minipage}{0.33\linewidth}
    \includegraphics[width=\linewidth]{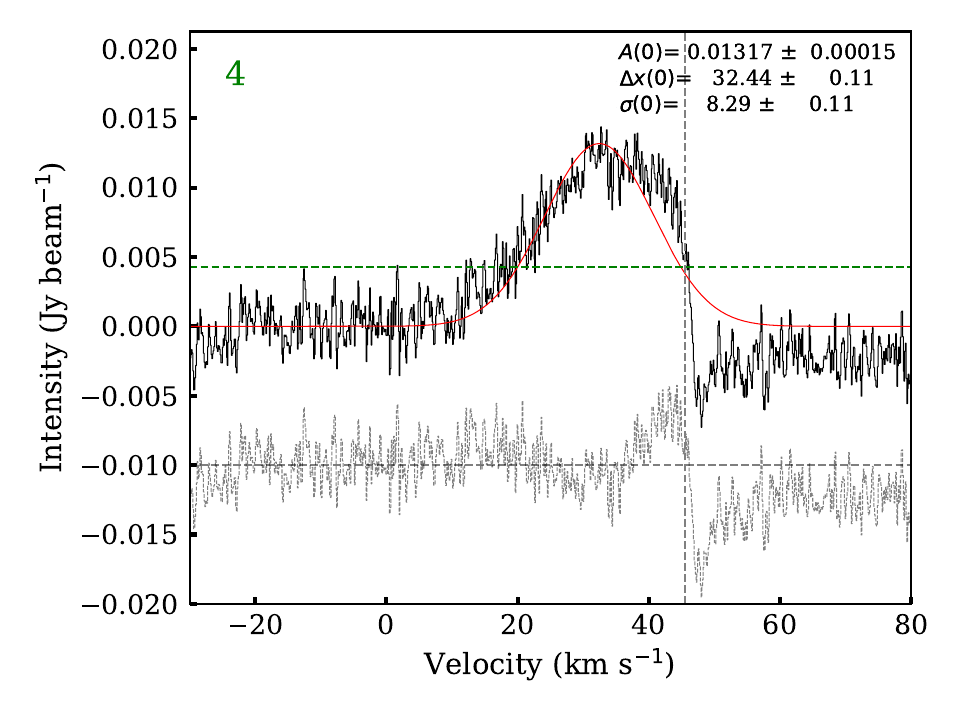}
\end{minipage}
\begin{minipage}{0.33\linewidth}
    \includegraphics[width=\linewidth]{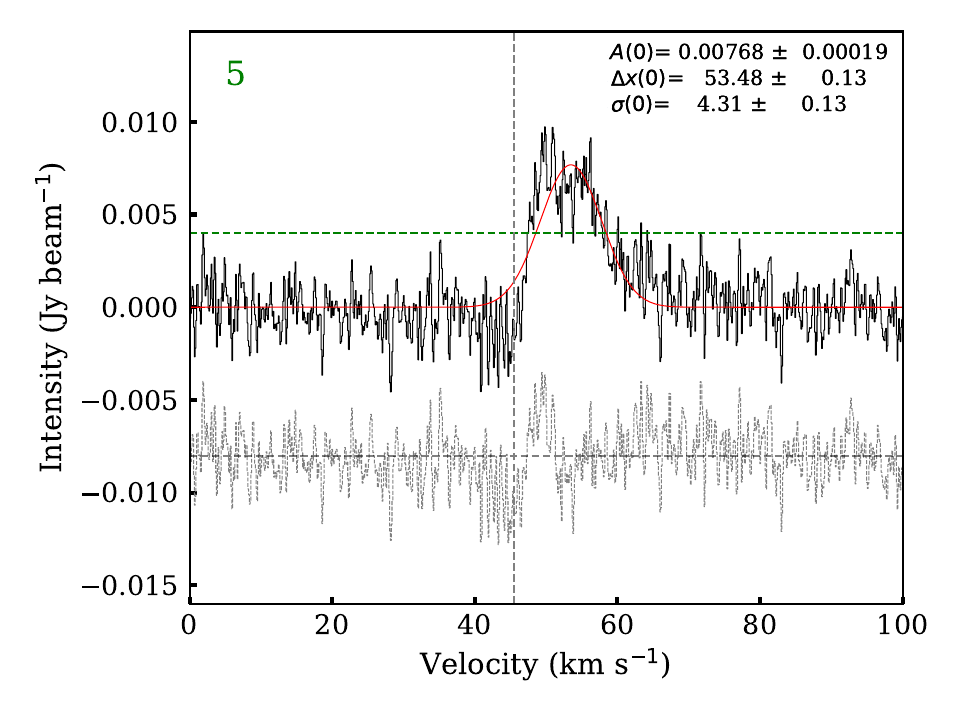}
\end{minipage}
\begin{minipage}{0.33\linewidth}
    \includegraphics[width=\linewidth]{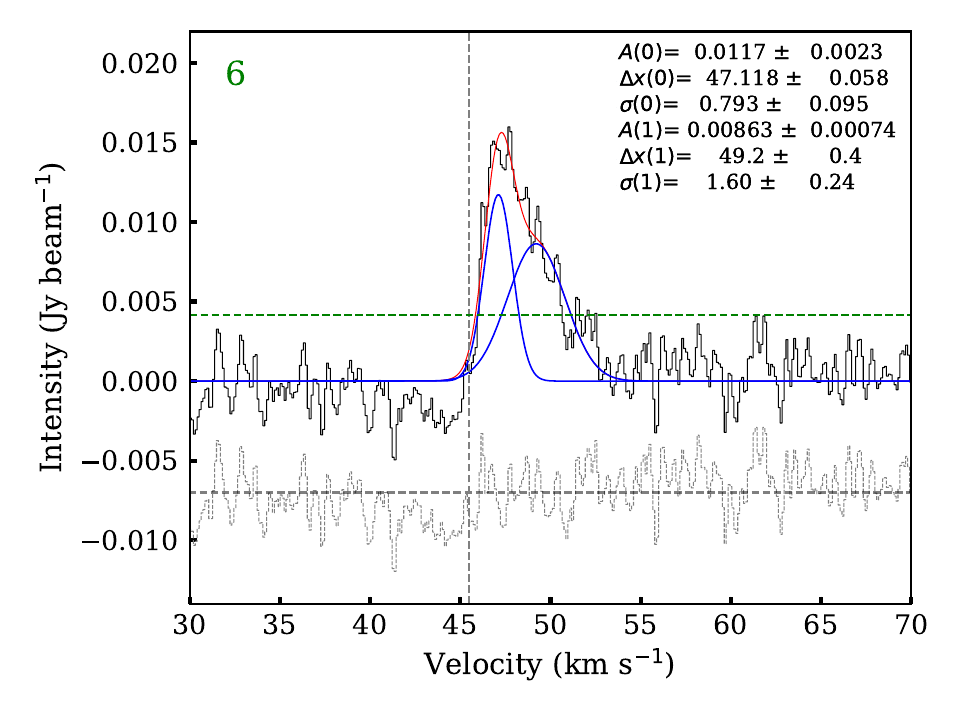}
\end{minipage}
\vspace{5pt} 
\begin{minipage}{0.33\linewidth}
    \includegraphics[width=\linewidth]{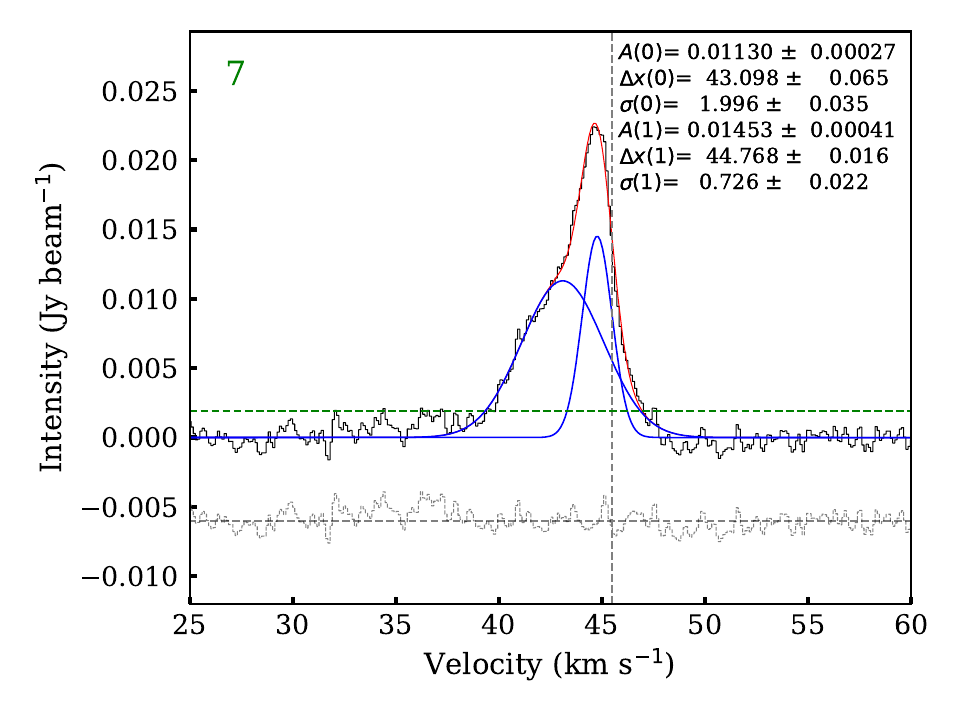}
\end{minipage}
\begin{minipage}{0.33\linewidth}
    \includegraphics[width=\linewidth]{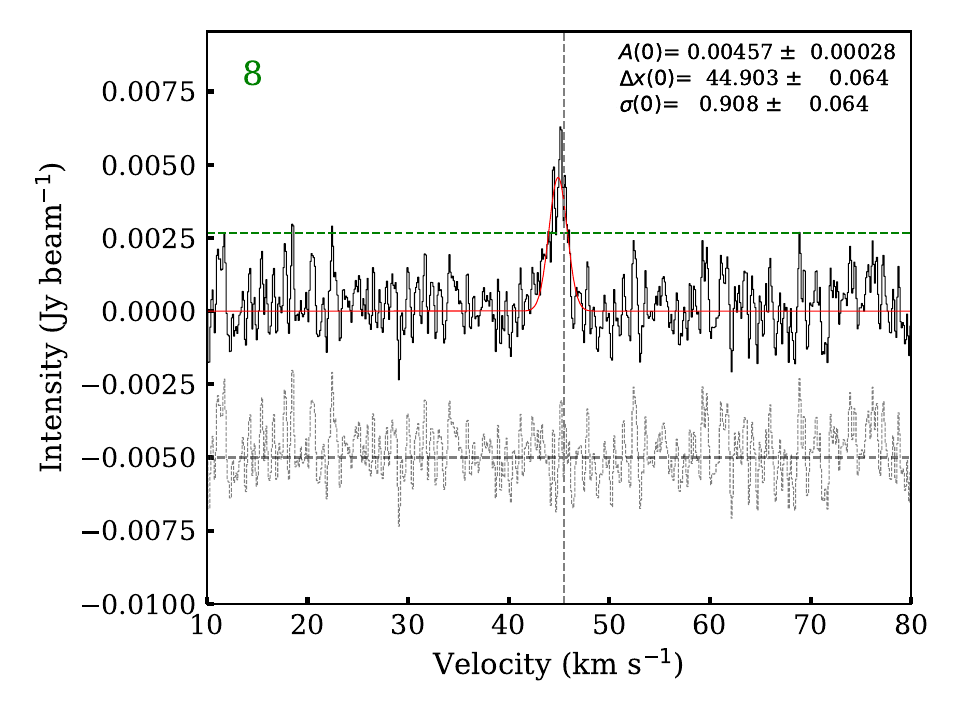}
\end{minipage}
\begin{minipage}{0.33\linewidth}
    \includegraphics[width=\linewidth]{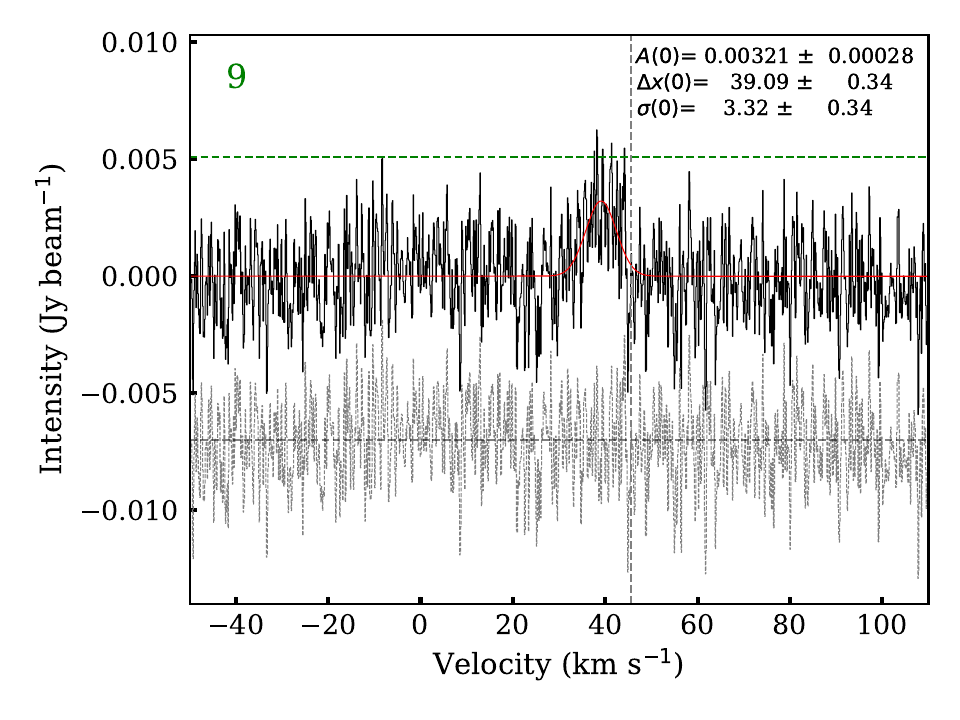}
\end{minipage}
\vspace{5pt} 
\begin{minipage}{0.33\linewidth}
    \includegraphics[width=\linewidth]{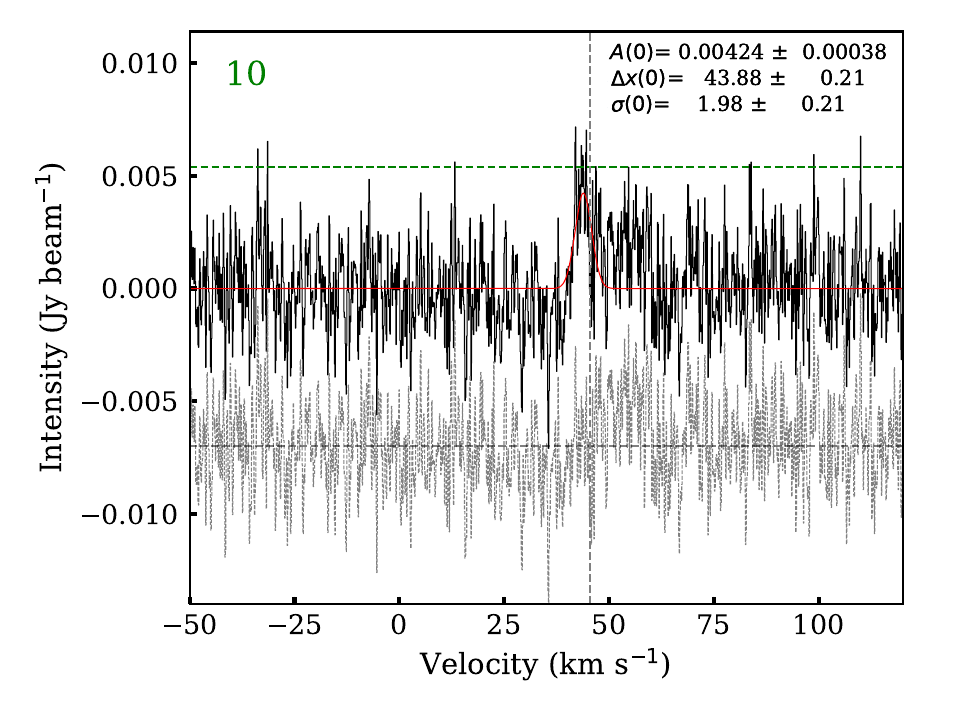}
\end{minipage}
\begin{minipage}{0.33\linewidth}
    \includegraphics[width=\linewidth]{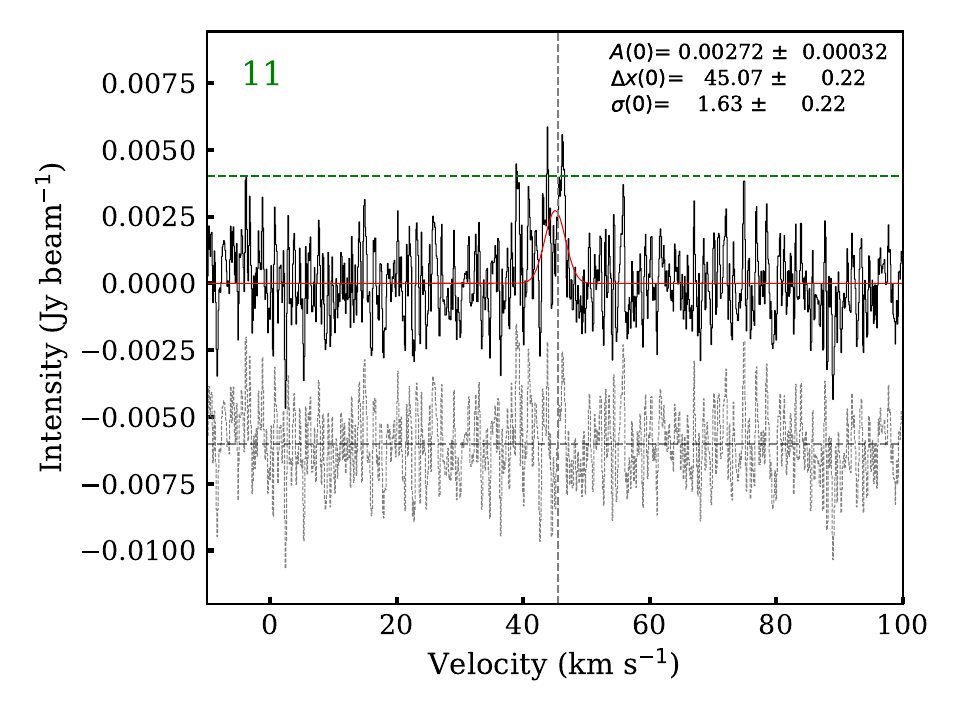}
\end{minipage}
\begin{minipage}{0.33\linewidth}
    \includegraphics[width=\linewidth]{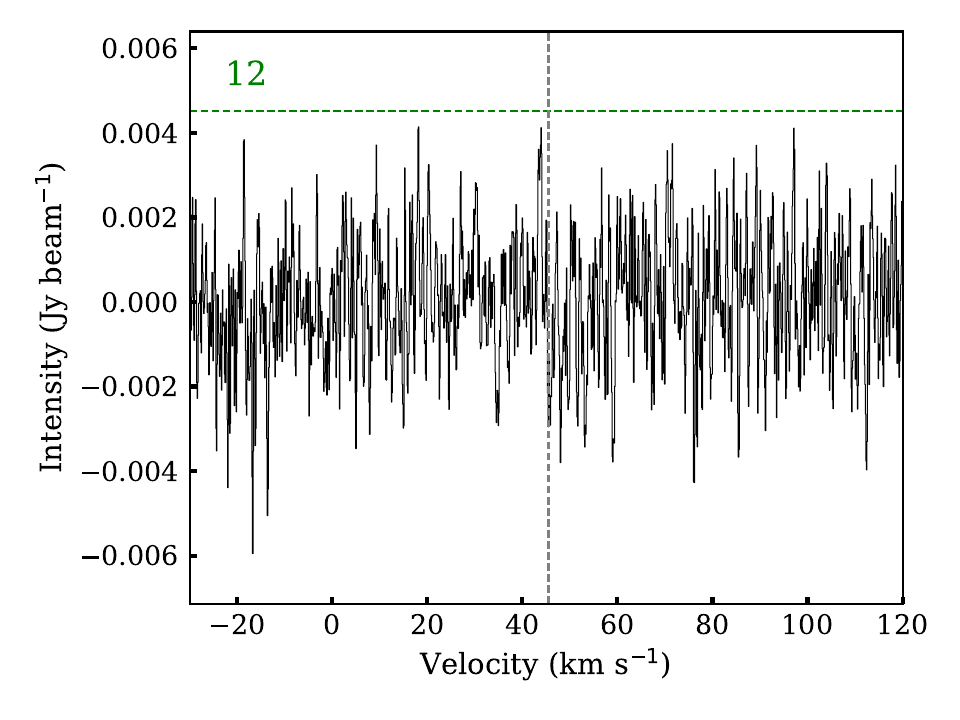}
\end{minipage}
\caption{Black curves represent the SiO spectra extracted from positions in Figure \ref{fig2}. The red curves report the total Gaussian fits to the SiO spectra using $\textit{pyspeckit}$ and the blue curves represent the Gaussian fit lines of the different components. The green dashed parallel line indicates the 3$\sigma$ noise levels of the SiO emission. The vertical dashed gray line marks the central velocity of Cloud H (45.5 km $^{-1}$). The residuals of the fitting results are shown as gray lines at the bottom of each panel, with the gray parallel line representing the residual zero baseline. The corresponding green numbers (see Figure \ref{fig2}) are displayed at the top left corner of each panel, and the Gaussian fitting results are shown at the top right corner.}
\label{figA2}
\end{figure*}

\begin{figure*}
\centering
\begin{minipage}{0.33\linewidth}
    \includegraphics[width=\linewidth]{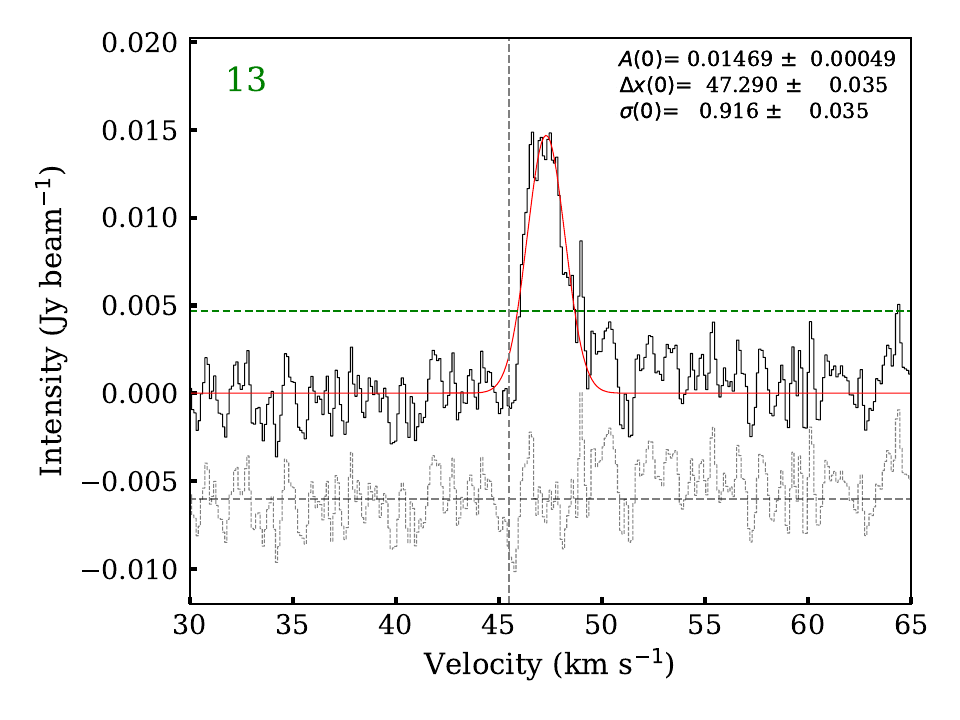}
\end{minipage}
\begin{minipage}{0.33\linewidth}
    \includegraphics[width=\linewidth]{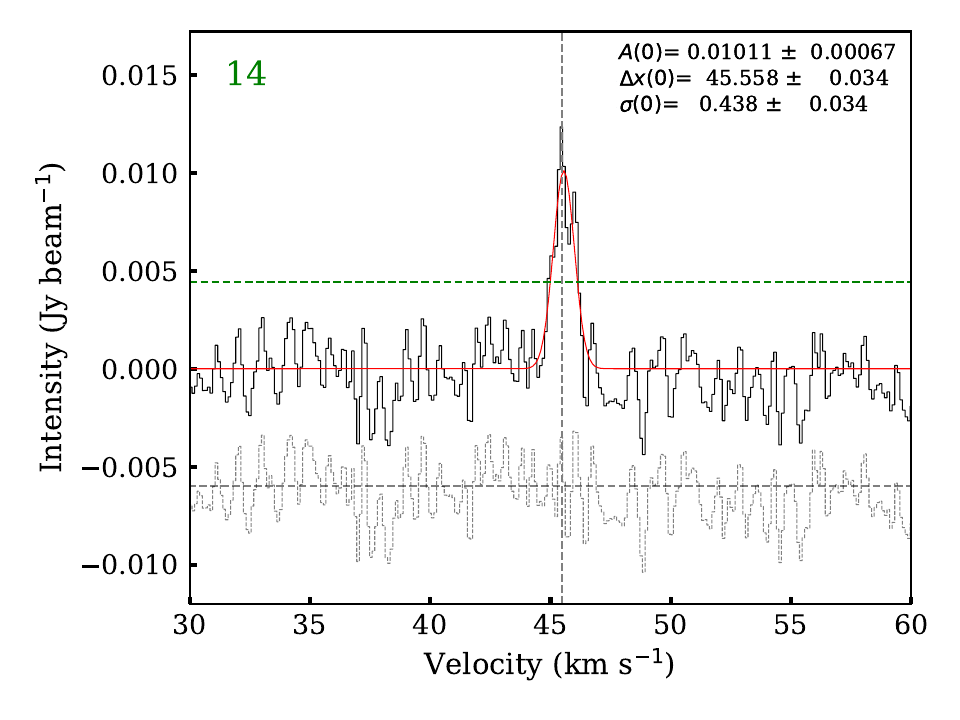}
\end{minipage}
\begin{minipage}{0.33\linewidth}
    \includegraphics[width=\linewidth]{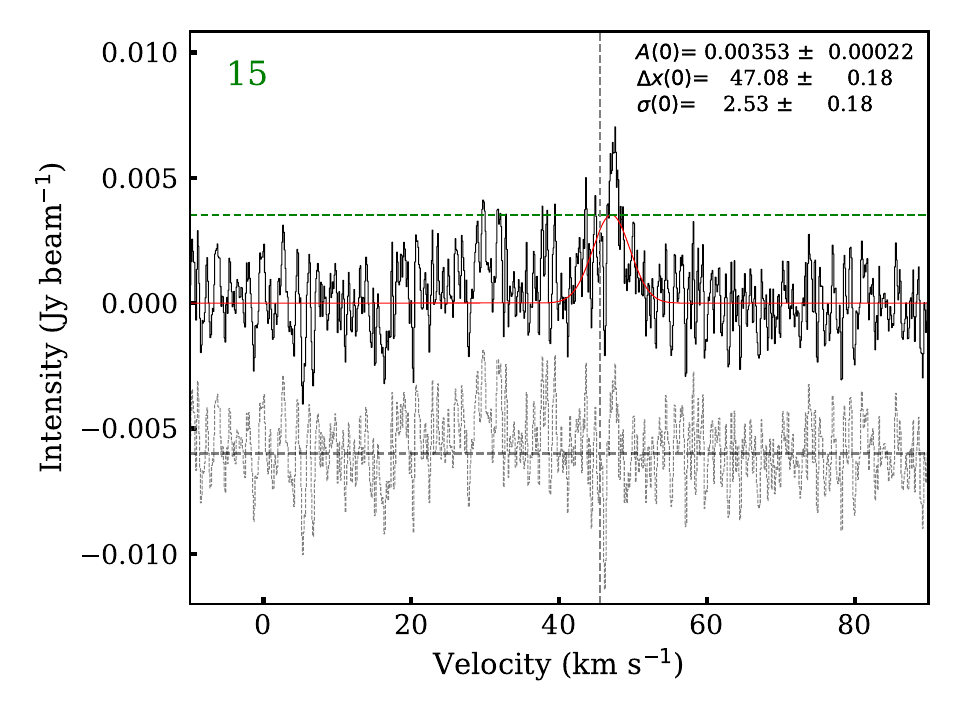}
\end{minipage}
\vspace{5pt} 
\begin{minipage}{0.33\linewidth}
    \includegraphics[width=\linewidth]{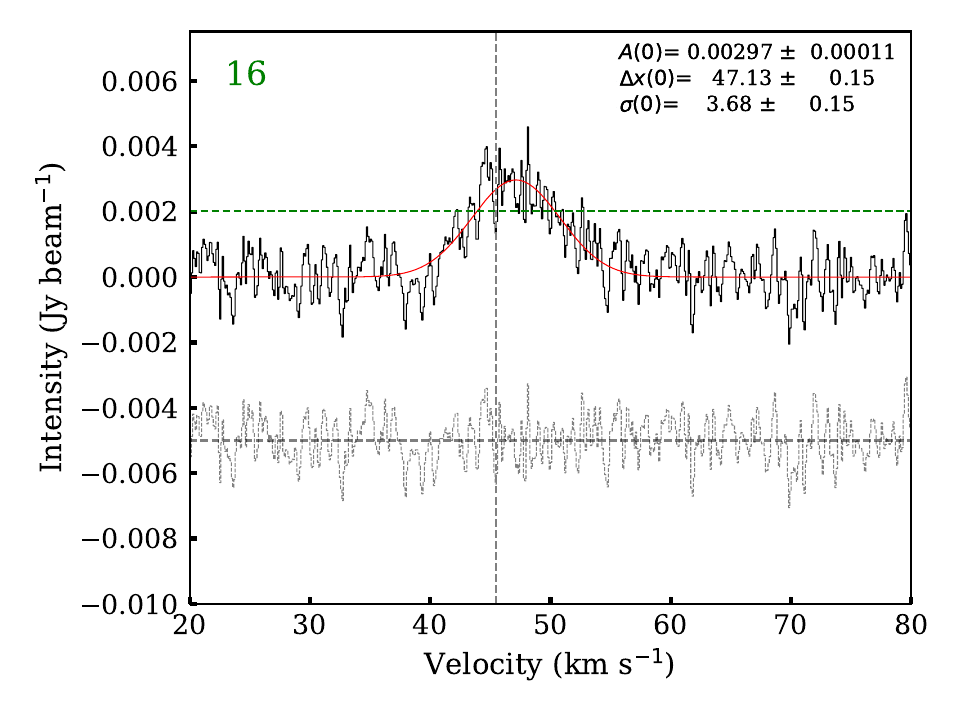}
\end{minipage}
\begin{minipage}{0.33\linewidth}
    \includegraphics[width=\linewidth]{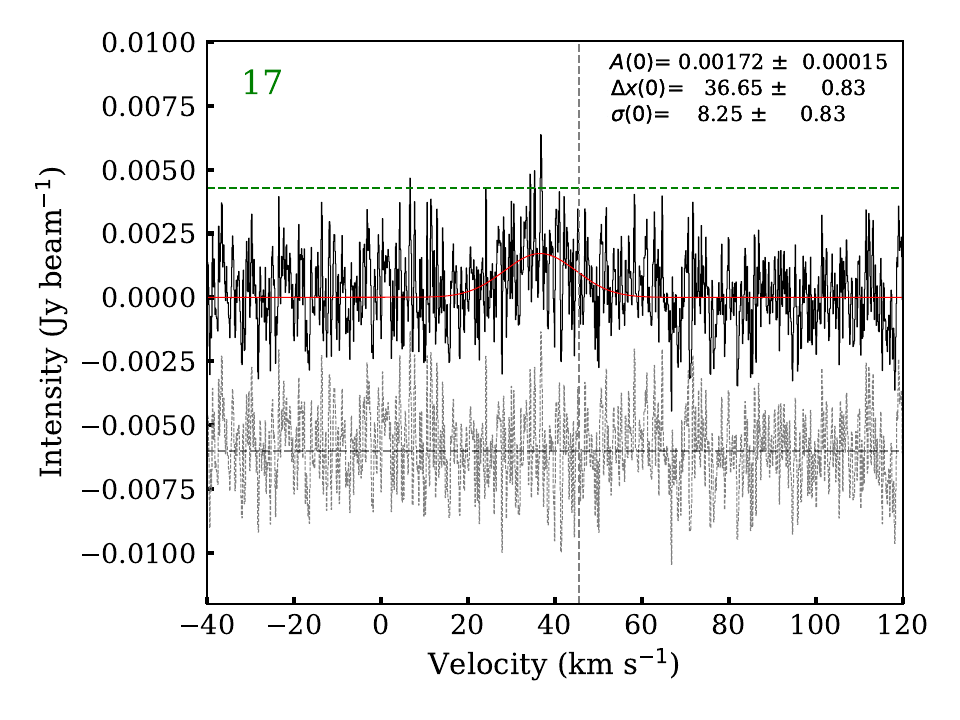}
\end{minipage}
\begin{minipage}{0.33\linewidth}
    \includegraphics[width=\linewidth]{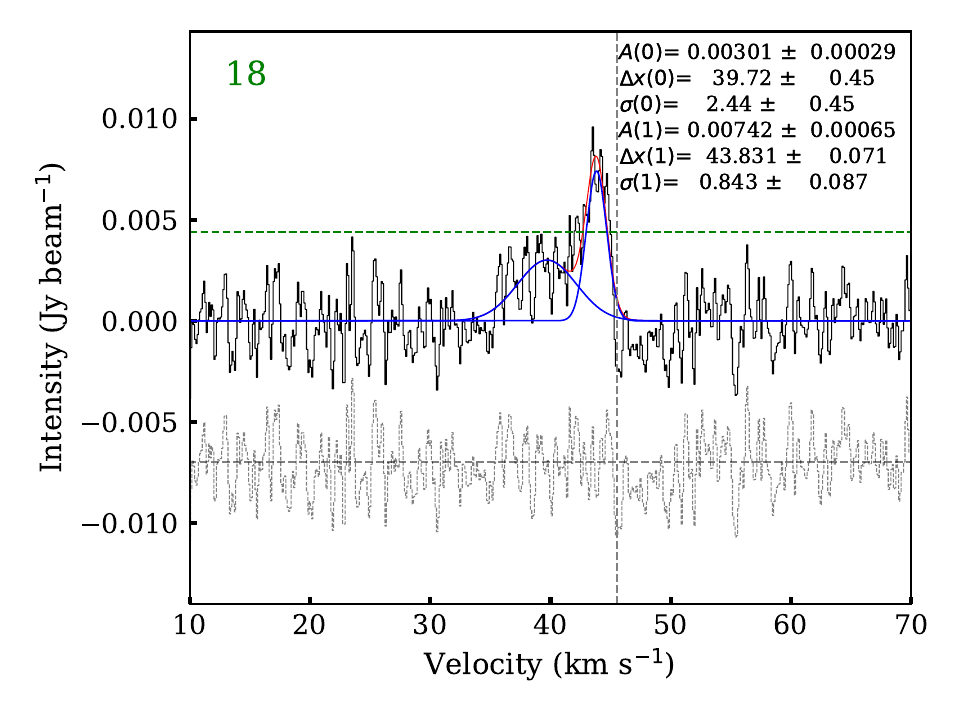}
\end{minipage}
\vspace{5pt} 
\begin{minipage}{0.33\linewidth}
    \includegraphics[width=\linewidth]{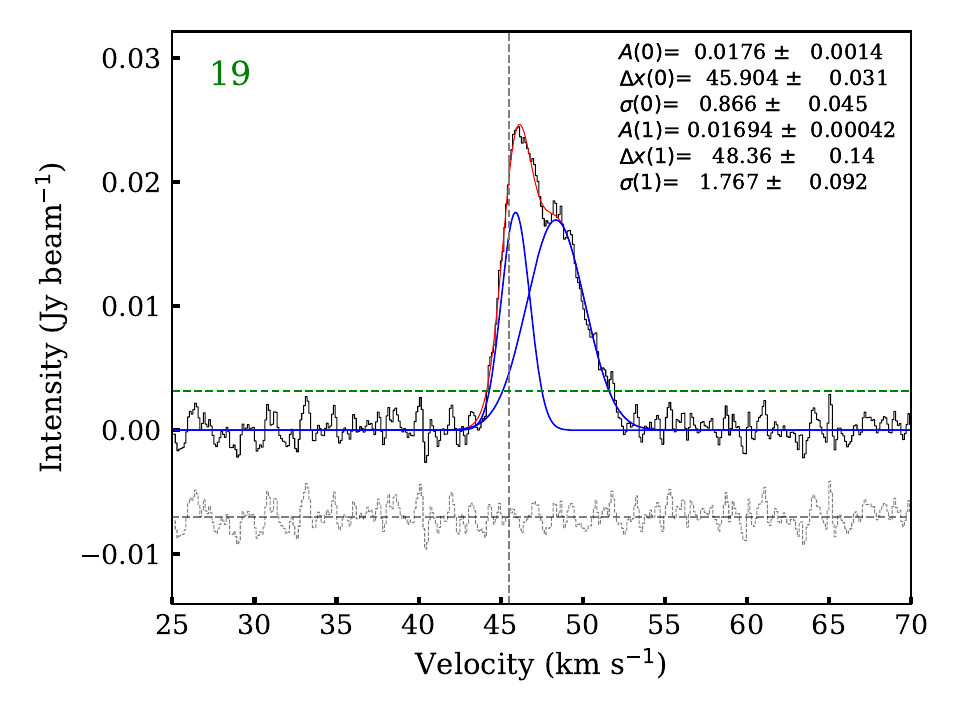}
\end{minipage}
\begin{minipage}{0.33\linewidth}
    \includegraphics[width=\linewidth]{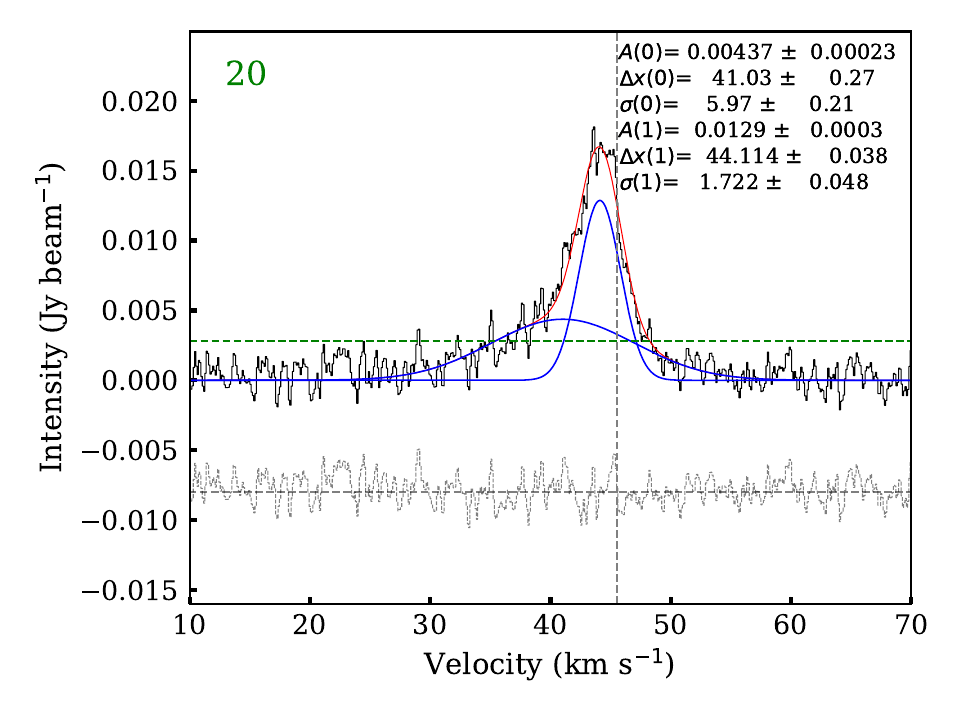}
\end{minipage}
\ContinuedFloat
\caption{Continued.}
\end{figure*}

\begin{figure*}
\centering
\begin{minipage}{0.33\linewidth}
    \includegraphics[width=\linewidth]{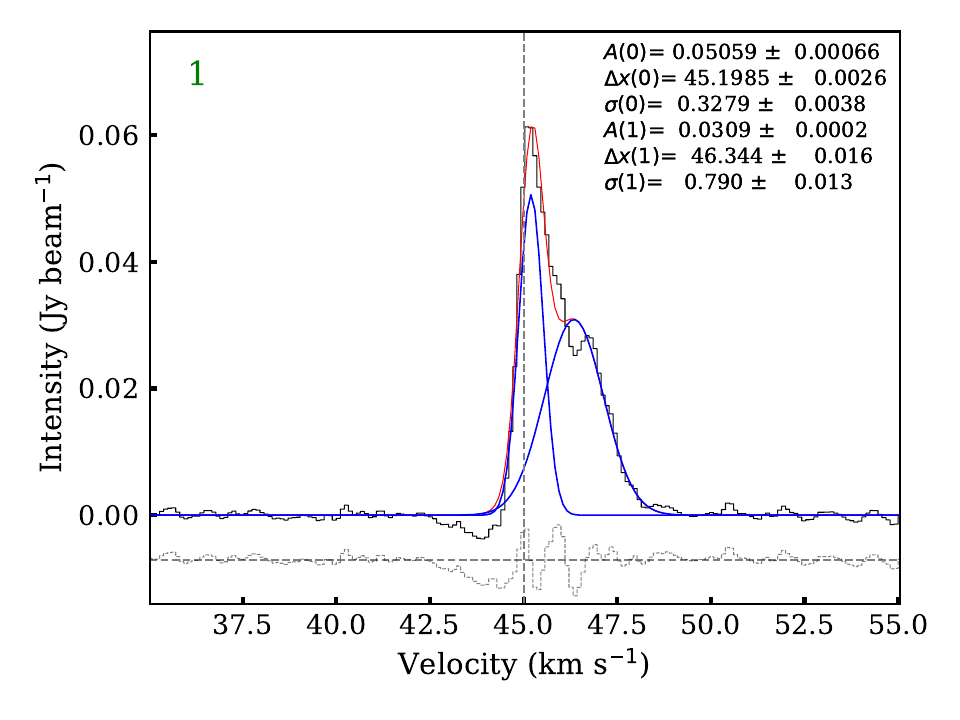}
\end{minipage}
\begin{minipage}{0.33\linewidth}
    \includegraphics[width=\linewidth]{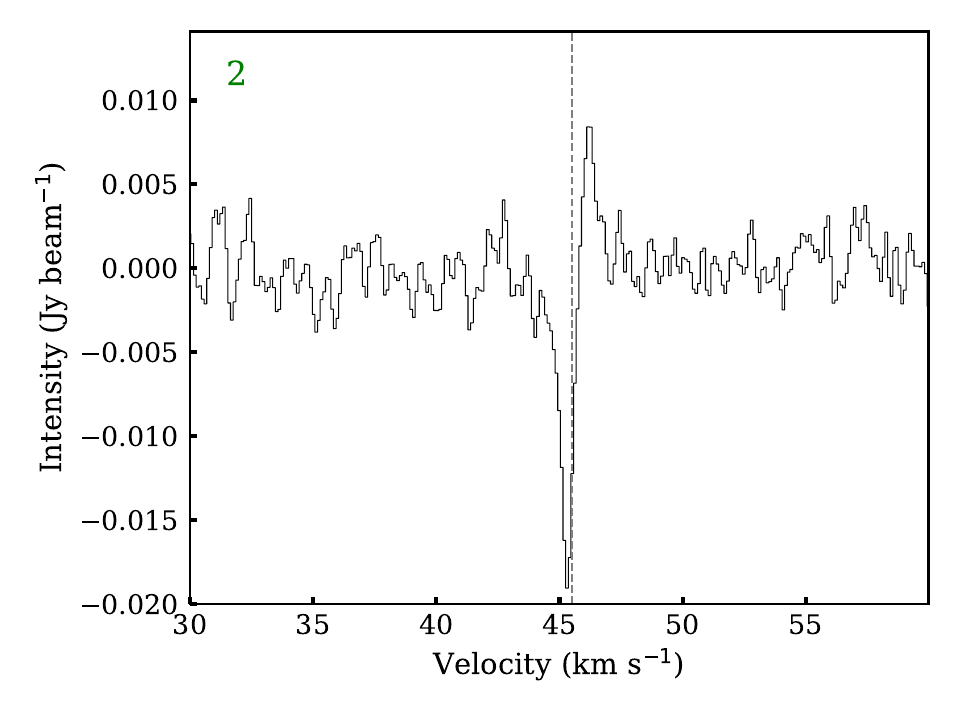}
\end{minipage}
\begin{minipage}{0.33\linewidth}
    \includegraphics[width=\linewidth]{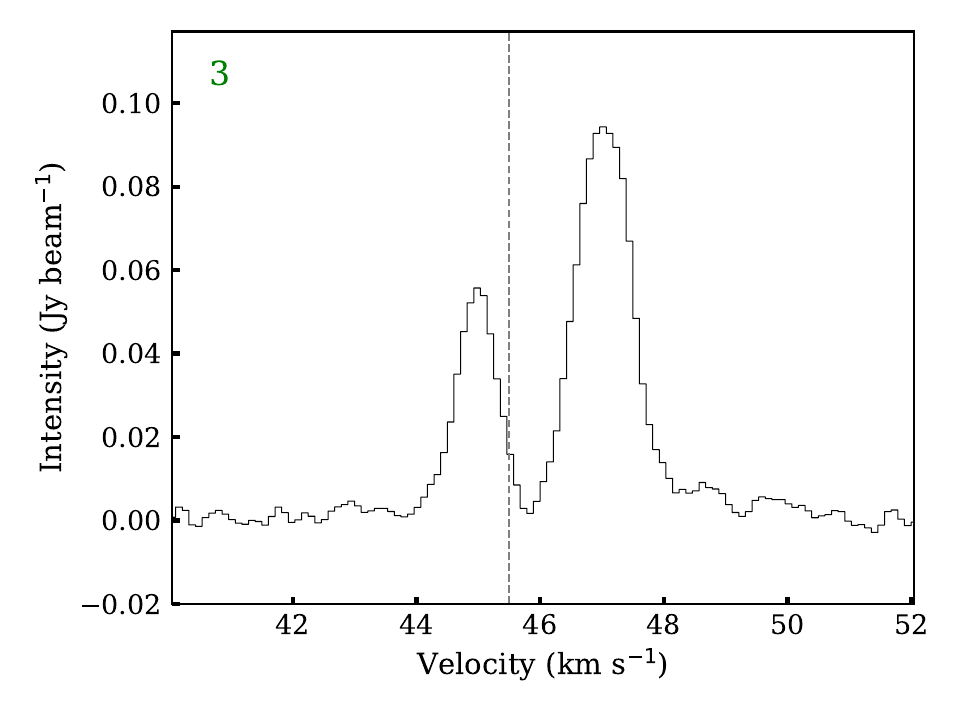}
\end{minipage}
\vspace{5pt}
\begin{minipage}{0.33\linewidth}
    \includegraphics[width=\linewidth]{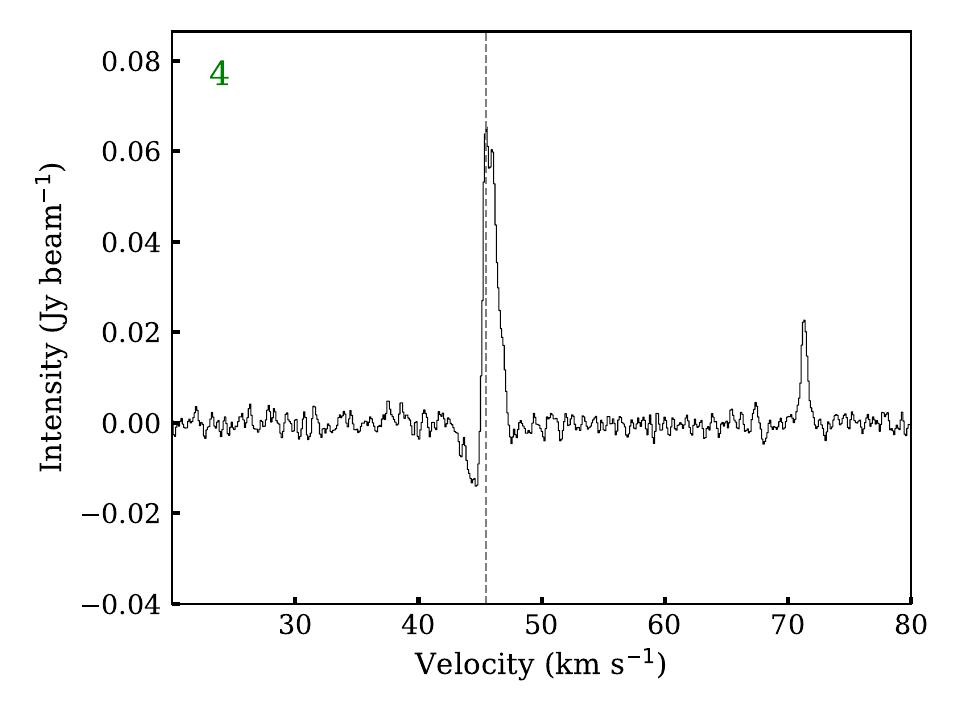}
\end{minipage}
\begin{minipage}{0.33\linewidth}
    \includegraphics[width=\linewidth]{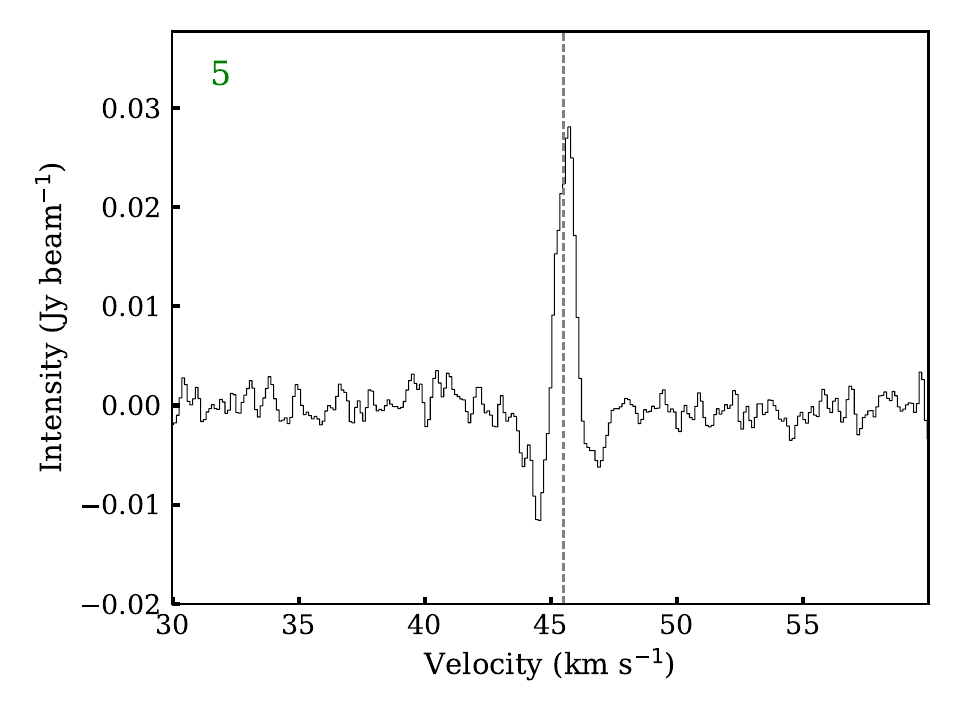}
\end{minipage}%
\begin{minipage}{0.33\linewidth}
    \includegraphics[width=\linewidth]{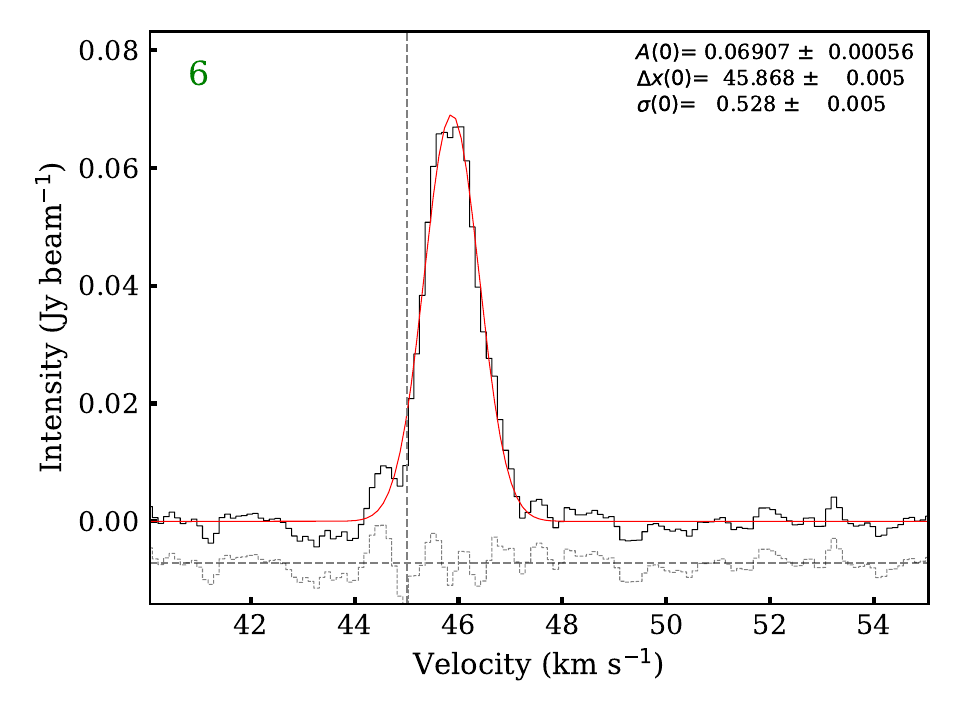}
\end{minipage}
\vspace{5pt}
\begin{minipage}{0.33\linewidth}
    \includegraphics[width=\linewidth]{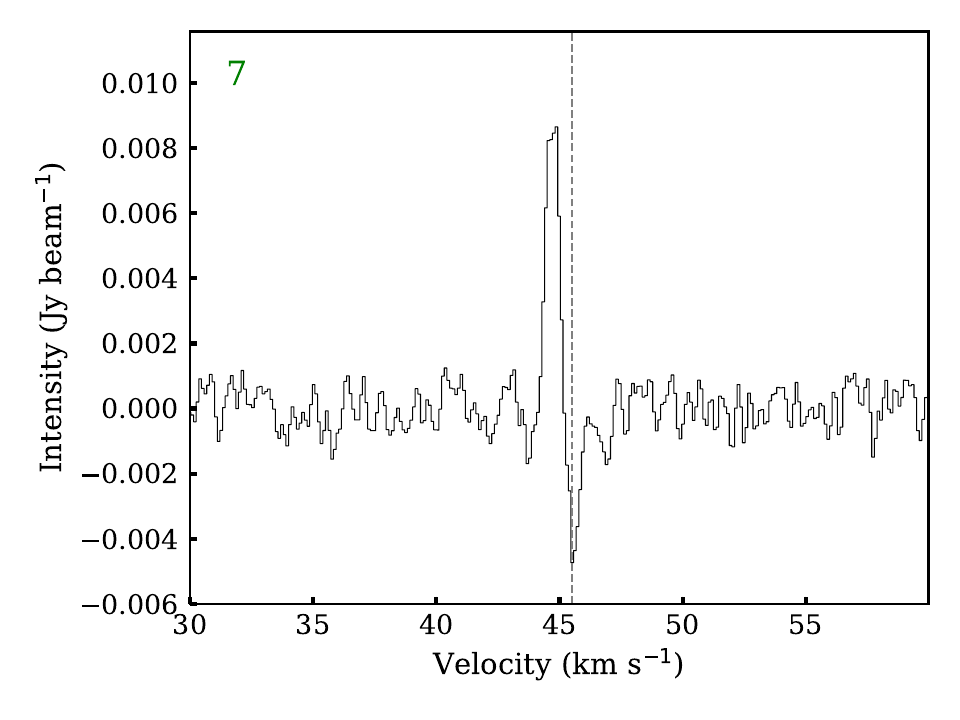}
\end{minipage}
\begin{minipage}{0.33\linewidth}
    \includegraphics[width=\linewidth]{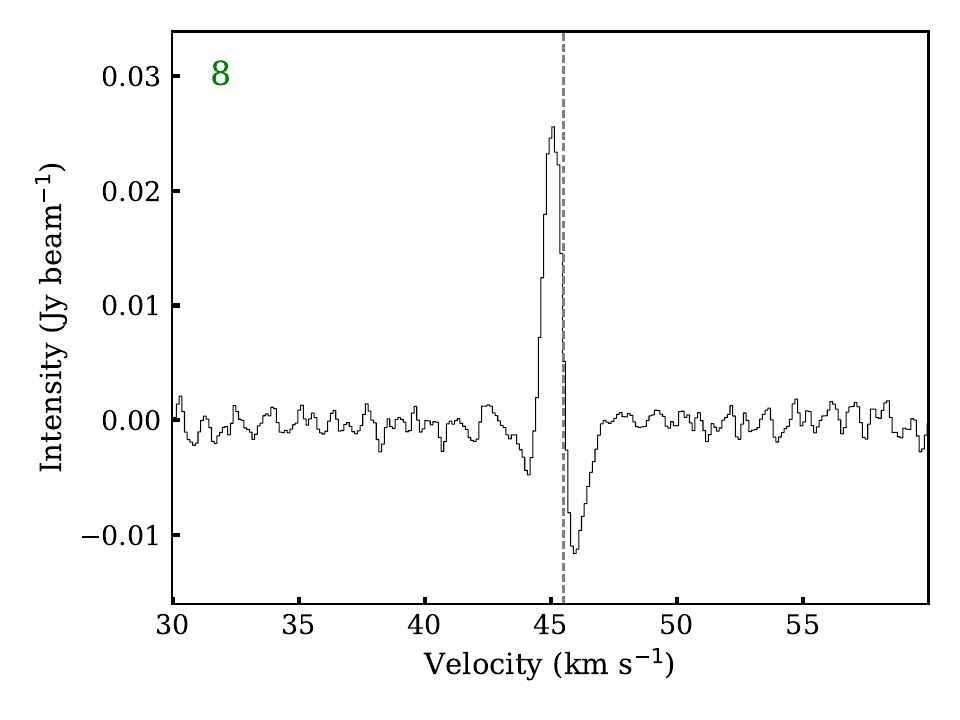}
\end{minipage}
\begin{minipage}{0.33\linewidth}
    \includegraphics[width=\linewidth]{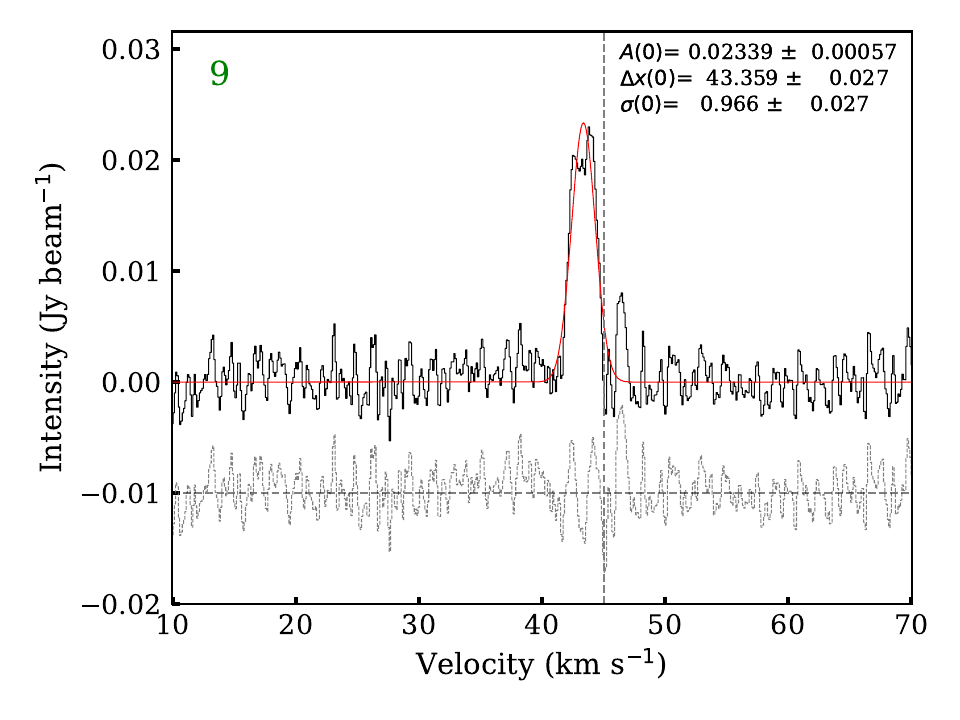}
\end{minipage}
\vspace{5pt}
\begin{minipage}{0.33\linewidth}
    \includegraphics[width=\linewidth]{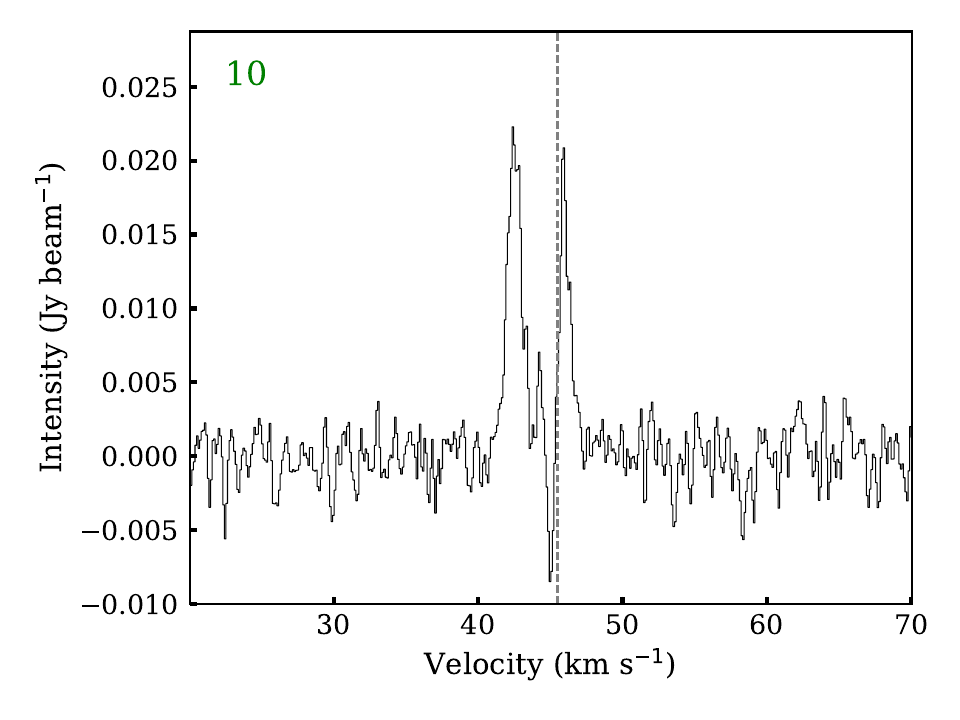}
\end{minipage}
\begin{minipage}{0.33\linewidth}
    \includegraphics[width=\linewidth]{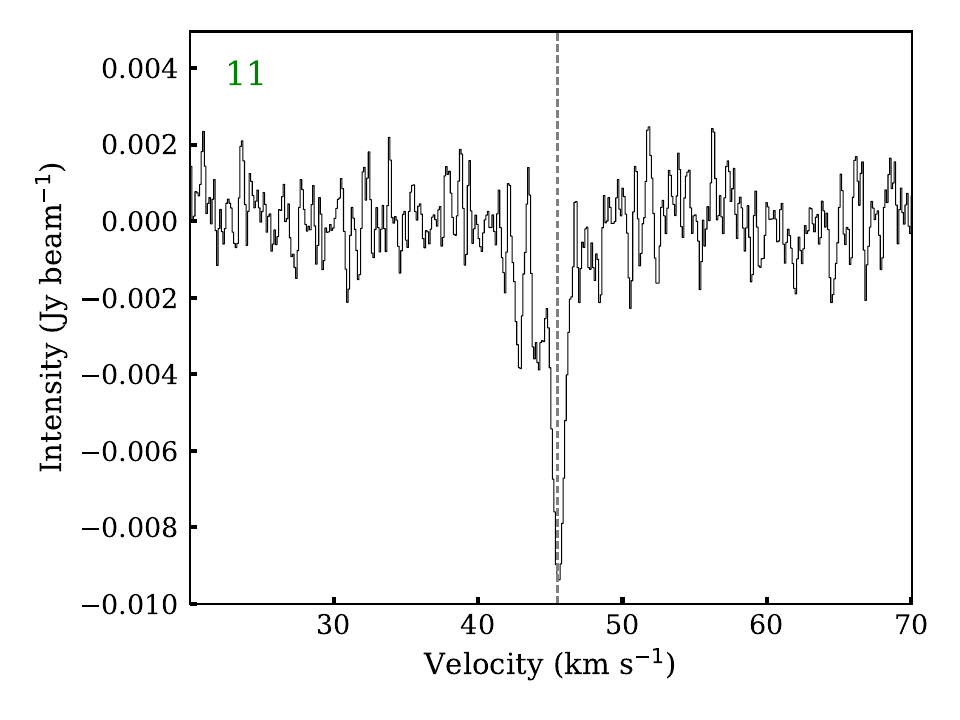}
\end{minipage}
\begin{minipage}{0.33\linewidth}
    \includegraphics[width=\linewidth]{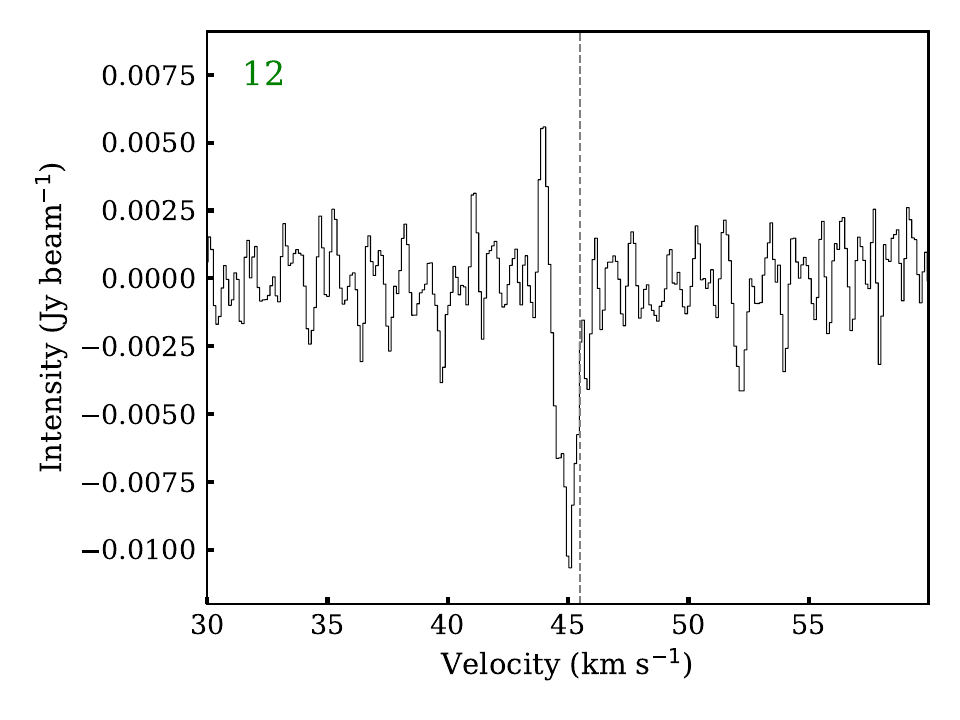}
\end{minipage}
\caption{Black curves represent the H$^{13}$CO$^+$ spectra extracted from positions in Figure \ref{fig2}. The vertical dashed gray line marks the central velocity of Cloud H (45.5 km $^{-1}$). The corresponding green numbers (see Figure \ref{fig2}) are displayed at the top left corner of each panel. Other symbols are the same
as in Figure \ref{figA2}.}
\label{figA3}
\end{figure*}

\begin{figure*}
\centering
\begin{minipage}{0.33\linewidth}
    \includegraphics[width=\linewidth]{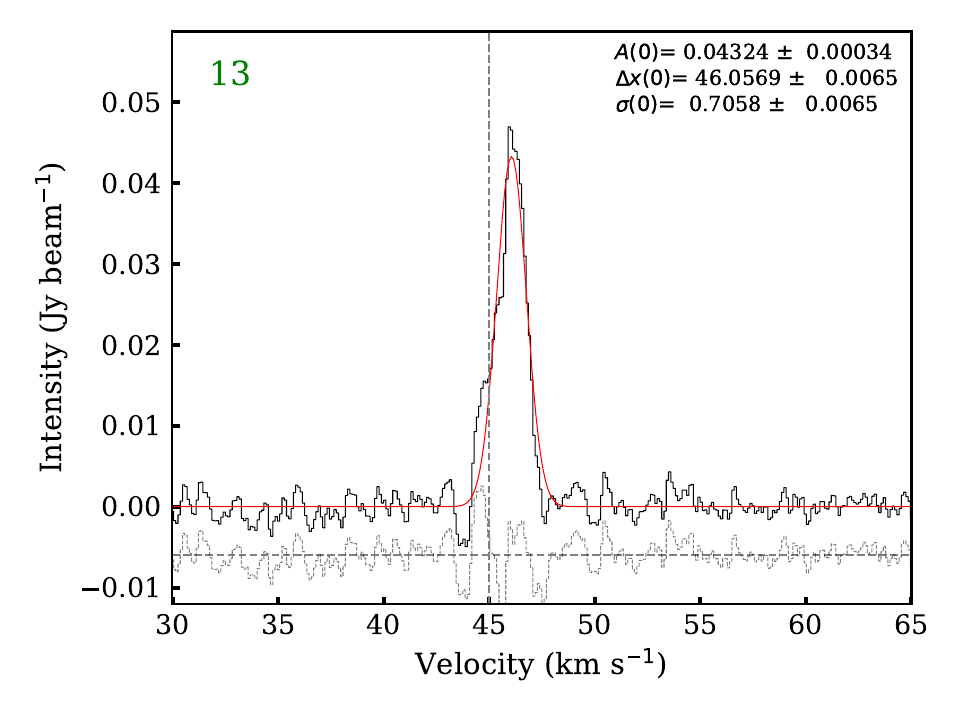}
\end{minipage}%
\begin{minipage}{0.33\linewidth}
    \includegraphics[width=\linewidth]{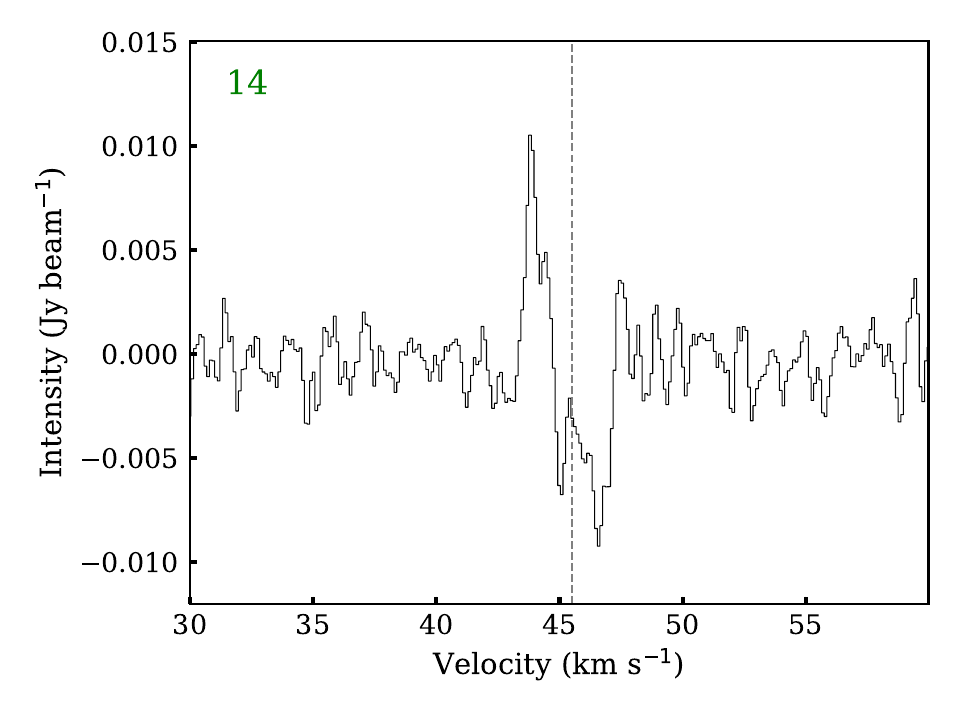}
\end{minipage}
\begin{minipage}{0.33\linewidth}
    \includegraphics[width=\linewidth]{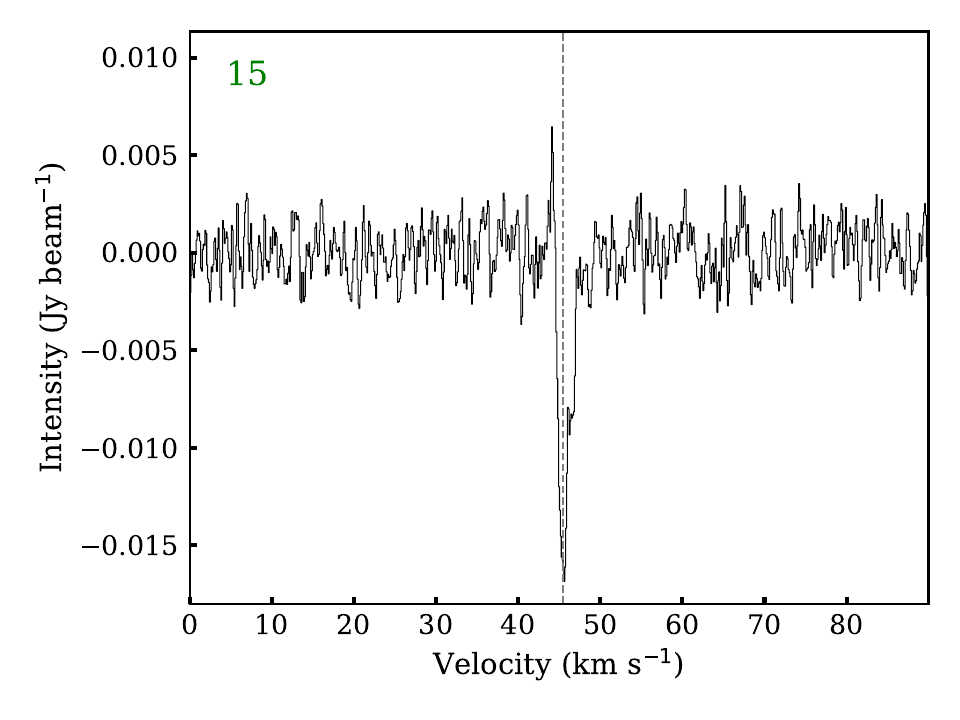}
\end{minipage}
\vspace{5pt} 
\begin{minipage}{0.33\linewidth}
    \includegraphics[width=\linewidth]{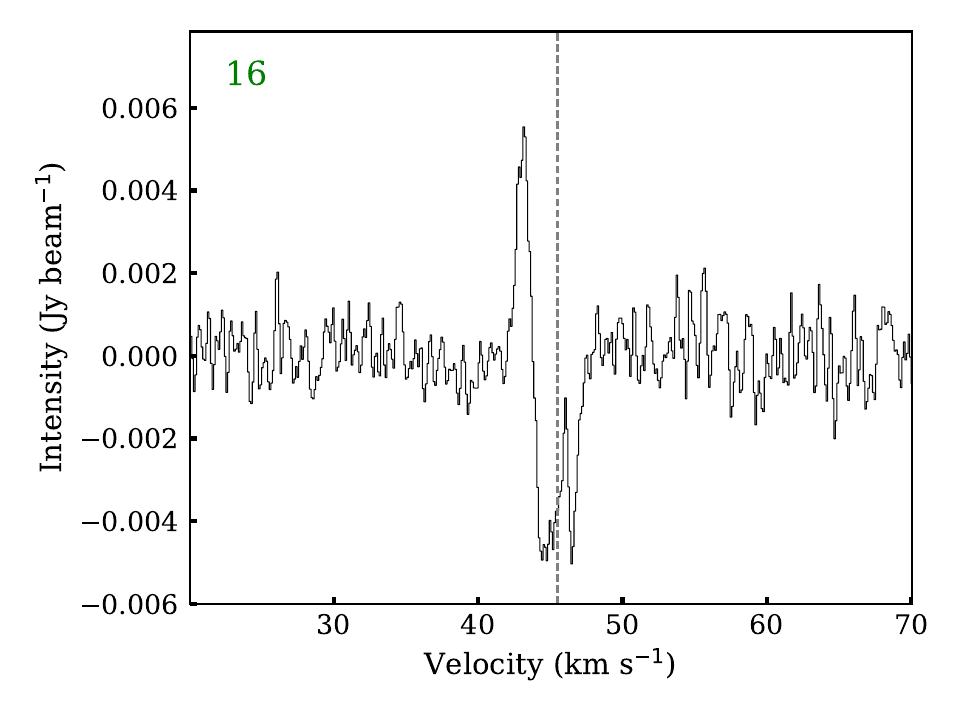}
\end{minipage}
\begin{minipage}{0.33\linewidth}
    \includegraphics[width=\linewidth]{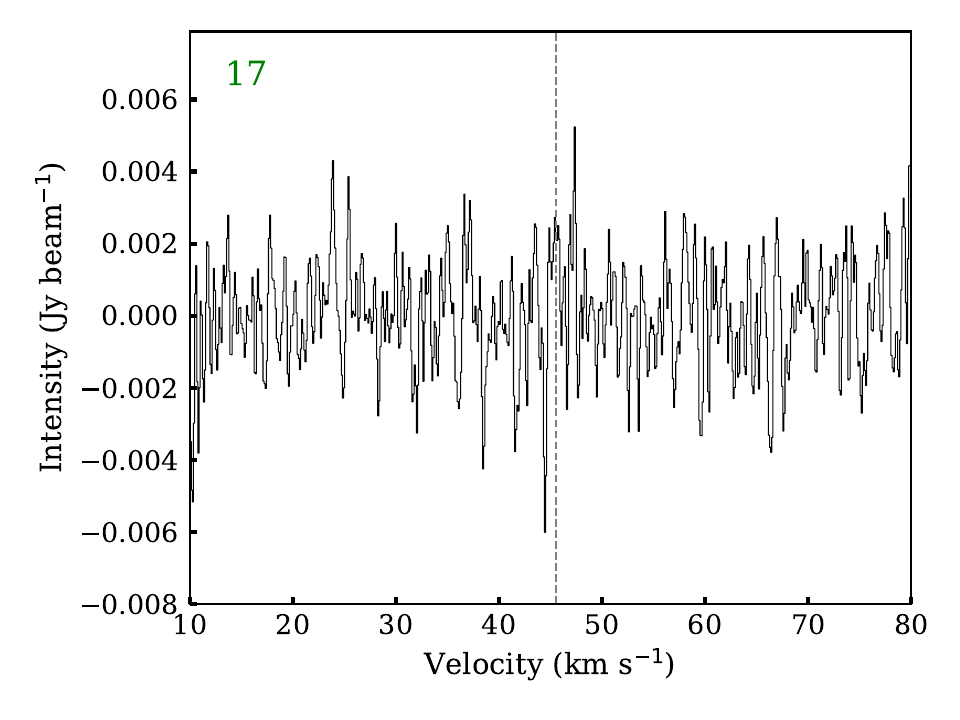}
\end{minipage}
\begin{minipage}{0.33\linewidth}
    \includegraphics[width=\linewidth]{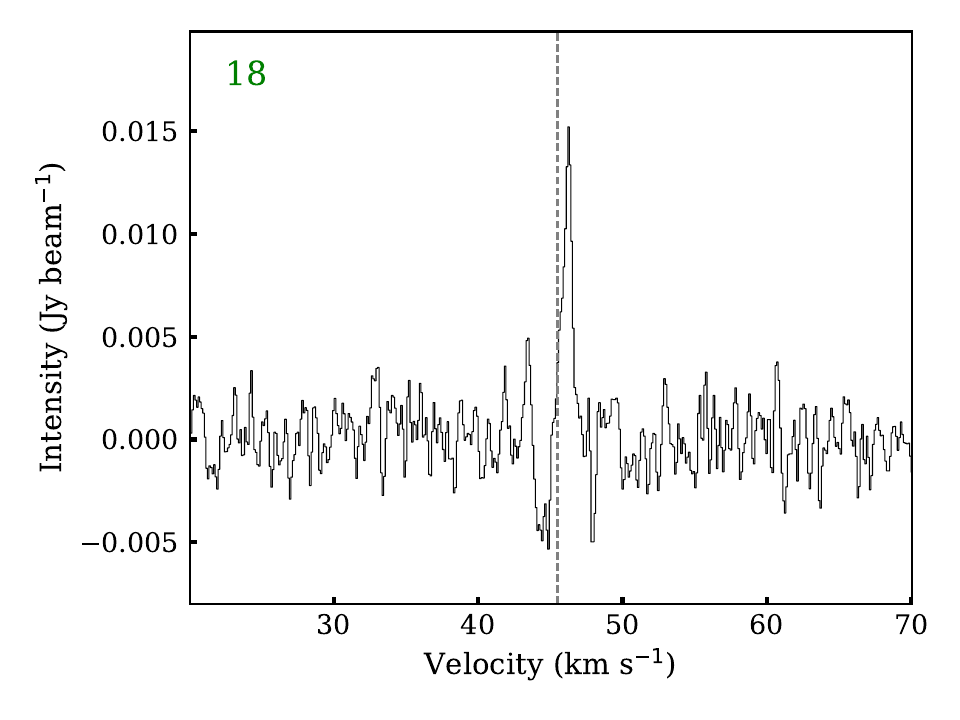}
\end{minipage}
\vspace{5pt}
\begin{minipage}{0.33\linewidth}
    \includegraphics[width=\linewidth]{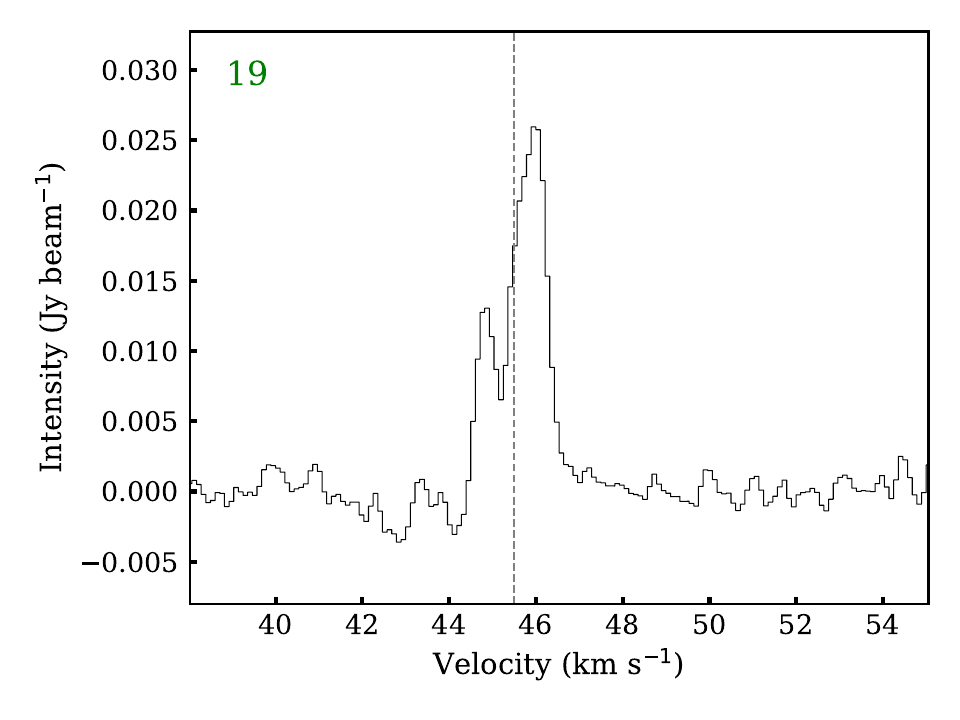}
\end{minipage}
\begin{minipage}{0.33\linewidth}
    \includegraphics[width=\linewidth]{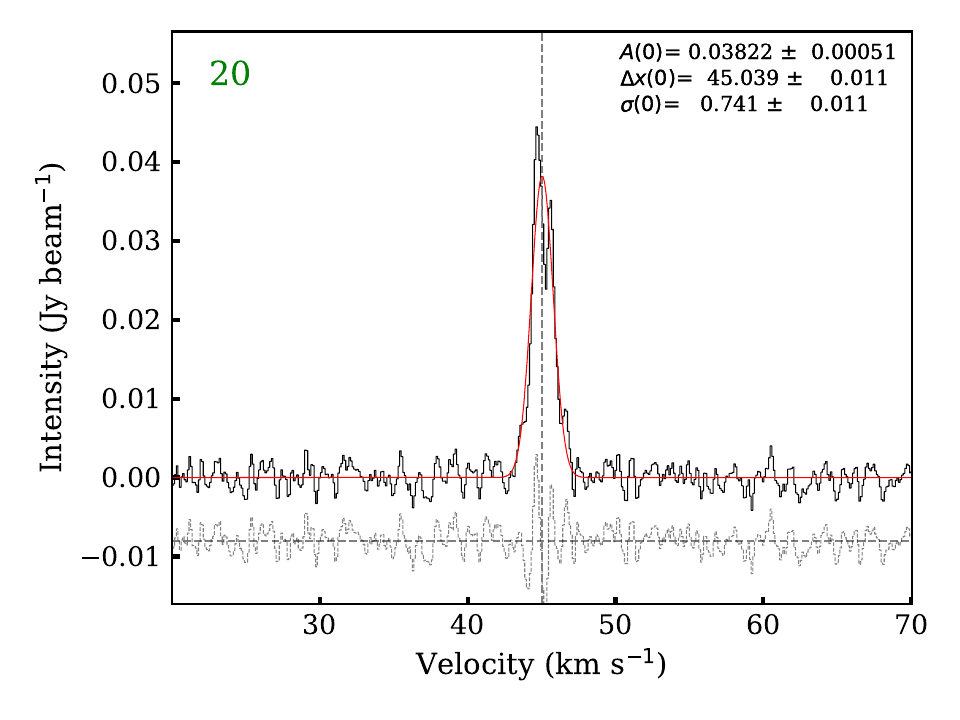}
\end{minipage}
\ContinuedFloat
\caption{Continued.}
\end{figure*}
\end{document}